\documentclass[11pt]{article}
\usepackage{jcapmod}

\usepackage{booktabs}
\usepackage[english]{babel}
\usepackage{amsmath,amssymb,amsbsy,amstext, amsthm, simplewick}
\usepackage{hyperref}
\usepackage{graphicx}
\usepackage{amsfonts}
\usepackage{amssymb}
\usepackage[small]{caption}
\usepackage{upgreek}
\usepackage{framed}
\usepackage{ulem}

\usepackage{colortbl}
\definecolor{lightgreen}{cmyk}{0.2, 0, 0.2, 0.2}
\definecolor{lightgray}{cmyk}{0.1,0.2,0,0.1}
\definecolor{lightgray2}{cmyk}{0.1,0.1,0,0.1}

\makeatletter
\newlength{\apb@width}
\newcommand{\autoparbox}[2][c]{\settowidth{\apb@width}{#2}\parbox[#1]{\apb@width}{#2}}

\makeatother

\numberwithin{equation}{section}

\def\beq{\begin{equation}}
\def\eeq{\end{equation}}
\def\be{\begin{eqnarray}}
\def\ee{\end{eqnarray}}
\def\beq{\begin{equation}}
\def\eeq{\end{equation}}
\def\bea{\begin{eqnarray}}
\def\eea{\end{eqnarray}}

\newcommand{\ex}[1]{\left\langle #1 \right\rangle}
\def\Beq{\begin{equation}\begin{aligned}}
\def\Eeq{\end{aligned}\end{equation}}

\def\eps{\varepsilon}

\def\dd{{\rm d}}

\def\e{\varepsilon}

\def\dd{{\rm d}}

\def\O{\mathcal{O}}

\def\deq{\hat{\mathcal{L}}}

\def\b{{\bf b}}

\def\Mpl{M_{\mathrm{Pl}}}

\def\k{{\bf k}}

\def\v{{\bf v}}
\def\x{{\bf x}}

\DeclareRobustCommand{\SkipTocEntry}[4]{}

\newcommand{\refeq}[1]{Eq.~(\ref{#1})}          
          
\newcommand{\reffig}[1]{Fig.~\ref{#1}}          
\newcommand{\refsec}[1]{Sec.~\ref{#1}}
\newcommand{\refss}[1]{Sec.~\ref{#1}}

\newcommand{\refapp}[1]{App.~\ref{#1}}

\newcommand{\reading}[1]{\hfill $ _{\text{\hyperref[#1]{ref}}} $}
\newcommand{\der}{\partial}
\newcommand{\expect}[1]{\langle #1 \rangle}
\newcommand{\intk}{\int_{\bf k}}

\renewcommand{\v}[1]{\ensuremath{\mathbf{#1}}} 

\newcommand{\then}{\quad \Rightarrow \quad }
\newcommand{\meff}{m_{\text{eff}}}

\begin{document}

\begin{titlepage}

\setcounter{page}{1} \baselineskip=15.5pt \thispagestyle{empty}

\bigskip\

\vspace{1cm}
\begin{center}

{\fontsize{20}{24}\selectfont  \sffamily \bfseries  The Conformal Limit of Inflation\\[0.5cm] in the Era of CMB Polarimetry}

\end{center}

\vspace{0.2cm}
\begin{center}
{\fontsize{13}{30}\selectfont  Enrico Pajer,$^{\clubsuit}$ Guilherme L. Pimentel,$^{\spadesuit, \hskip 1pt\bigstar}$ and Jaap V. S. van Wijck$^{\clubsuit}$} 
\end{center}

\begin{center}

\vskip 8pt
\textsl{$^{\clubsuit}$ Institute for Theoretical Physics and Center for Extreme Matter and Emergent Phenomena,\\ Utrecht University, Princetonplein 5, 3584 CC Utrecht, The Netherlands}
\vskip 8pt
\textsl{$^{\spadesuit}$ Institute of Physics, Universiteit van Amsterdam,\\ Science Park, Amsterdam, 1090 GL, The Netherlands}  
\vskip 8pt
\textsl{$^{\bigstar}$ Department of Applied Mathematics and Theoretical Physics,\\ 
Cambridge University, Cambridge, CB3 0WA, UK}

\vskip 7pt

\end{center}

\vspace{1.2cm}
\hrule \vspace{0.3cm}
\noindent {\sffamily \bfseries Abstract} \\[0.1cm]
We argue that the non-detection of primordial tensor modes has taught us a great deal about the primordial universe. In single-field slow-roll inflation, the current upper bound on the tensor-to-scalar ratio, $ r<0.07 $ ($95\% $ CL), implies that the Hubble slow-roll parameters obey $ \e\ll\eta $, and therefore establishes the existence of a new hierarchy. We dub this regime the conformal limit of (slow-roll) inflation, and show that it includes Starobinsky-like inflation as well as all viable single-field models with a sub-Planckian field excursion. In this limit, all primordial correlators are constrained by the full conformal group to leading non-trivial order in slow-roll. This fixes the power spectrum and the full bispectrum, and leads to the ``conformal'' shape of non-Gaussianity. The size of non-Gaussianity is related to the running of the spectral index by a consistency condition, and therefore it is expected to be small. In passing, we clarify the role of boundary terms in the $ \zeta $ action, the order to which constraint equations need to be solved, and re-derive our results using the Wheeler-deWitt formalism.
\vskip 10pt
\hrule
\vskip 10pt

\vspace{0.6cm}
 \end{titlepage}

\tableofcontents


\section{Introduction and Summary}

\subsection{Introduction}

Current cosmological observations are well-described by the $\Lambda$CDM model. This model assumes an almost scale-invariant initial power spectrum of fluctuations in a specific scalar mode, known as the adiabatic mode. These initial conditions are elegantly derived from inflation, where the (approximate) scale invariance follows from one of the isometries of (quasi) de Sitter spacetime. One of the simplest realizations of this scenario is a single scalar field that slowly rolls towards the bottom of a potential, thus providing a physical clock for the the inflationary period. This particular model of slow-roll inflation has indeed received a great deal of attention. 

An exciting signature of inflation are primordial tensor modes. However, despite much observational effort, a detection of primordial tensor perturbations has so far proven elusive, and current bounds for the tensor-to-scalar ratio $r$ give $ r<0.07 $ ($ 95\% $ CL) \cite{Array:2015xqh}. It is important to ask what we learn from this bound and from the many future improvements thereof. Naively, besides the exclusion of a handful of models, the \textit{absence} of a detection of $ r $ gives us little new information about the early universe.

In this paper, we point out that, within single-field, slow-roll, canonical models, the non-detection of $ r $ actually teaches us a great deal. The experimental detection of a deviation from the Harrison-Zeldovich spectrum, together with the current lower bound for the amplitude of tensor modes, implies a hierarchy between the slow-roll parameters that characterize the time dependence of the Hubble parameter during inflation. A consequence of this hierarchy is that all primordial correlators are constrained by conformal symmetry.  

Using this hierarchy, which we dub the {\it conformal limit of inflation}, we can determine completely the shape of the power spectrum and bispectrum of scalar fluctuations. Our method combines conformal symmetry and a consistency condition for the squeezed limit of the bispectrum. The bispectrum contains a local shape and a ``conformal" shape that has been studied in a different context in the literature~\cite{Zaldarriaga:2003my,Seery:2008qj,Creminelli:2011mw,Kundu:2014gxa}. The amplitude of this bispectrum is small, as the local shape is parametrically of the size of the scalar tilt, while the conformal shape is of the size of the running of the tilt.

One nice implication of this result is that the overall size of the bispectrum is tied to observables in the scalar power spectrum, namely the tilt and its running. Hence, a new single-field consistency condition emerges in the scalar sector, which is in principle testable even for negligible tensor modes. If future experiments keep pushing the upper bounds on $r$ and $f_{\rm NL}$, this consistency condition may turn out to be the ultimate test of the simplest single-field inflationary models.

Below we summarize our main results more quantitatively, while outlining the various sections of the paper.

\subsection{Summary of Results}
The simplest model of inflation consists of a single, canonical, minimally coupled scalar field. Assuming that gravity is described by General Relativity (GR), one is lead to consider the action
\bea\label{equ:infac}
S=\int \dd^4x\sqrt{-g} \left[ \frac{\Mpl^2}{2} R-\frac{(\nabla\phi)^2}{2}-V(\phi)\right]\,.
\eea
We will focus exclusively on this action and postpone further comments on more general single-clock models to the discussion, in \refsec{disc}. The background solution of \eqref{equ:infac} is a FLRW spacetime with Hubble parameter $H(t)$ which is approximately constant in the slow-roll regime. Small time dependence is parametrized by the two slow-roll parameters $\varepsilon\equiv -\dot H/H^2$ and $\eta\equiv \dot\varepsilon/(\varepsilon H)$. Within the model \refeq{equ:infac}, in the first order slow-roll approximation, one finds the well-known expressions
\be\label{srns}
1-n_{s}=2\e+\eta\,,\quad r=16\e\,,
\ee
where $ n_{s} $ is the scalar spectral tilt and $ r $ is the tensor-to-scalar ratio. Since observations tell us that \cite{Ade:2015lrj,Array:2015xqh}
\be\label{data}
1-n_{s}=0.0355 \pm0.005 \,  \text{(68\% CL)}\quad \text{and}\quad r<0.07 \, \text{(95\% CL)}\,,
\ee
we conclude that 
\be\label{hiercl}
\boxed{\e<0.0044< \frac{\eta}{6}\ll \eta\,.}
\ee
In words, \textit{since we have observed a non-vanishing scalar spectral index but no primordial tensor modes, we have discovered a new hierarchy of the slow-roll parameters, namely $ \e\ll \eta $}.\footnote{The generalization to $ c_{s}\neq1  $ is discussed in \refsec{disc}.} As CMB polarization experiments forge ahead in their quest for primordial tensor modes, in the event of no detection, the upper bound on $ r $ gets stronger and this hierarchy becomes more and more pronounced. Since $ \e $ and $ \eta $ have traditionally been treated on the same footing, some of the discussion of slow-roll inflation might need to be updated in light of this new hierarchy.

We call the model in \refeq{equ:infac} in the regime $ \e\ll\eta $ the \textit{conformal limit of inflation}. This limit is spelled out in detail in \refss{ss:conflim}. The conformal limit describes all viable, single-field, slow-roll canonical models. For example, small-field models with a subPlanckian inflaton displacement, $ \Delta\phi\ll \Mpl $ are a specific case of \refeq{hiercl}. In fact, the Lyth bound \cite{Lyth:1996im} implies
\be\label{smallfield}
\e\lesssim \left( \frac{\Delta \phi}{\Mpl} \right)^{2}\frac{1}{N^{2}}\ll \frac{1}{N^{2}}\lesssim \eta^{2}\,,
\ee 
where $ N\sim 60 $ is the duration of the observable part of inflation, $ \e $ and $ \eta $ are evaluated at CMB scales and the last inequality follows from $  \eta\simeq 1-n_{s}\simeq 0.035 > 1/60  $. Certain models with Planckian field excursion, such as Starobinsky-like inflation, are also particular cases of the conformal limit (see \refsec{ssec:staro}).

One remarkable fact about the conformal limit, which according to \refeq{hiercl} might well describe our universe, is that \textit{all primordial correlators are constrained by conformal symmetry}, up to small, slow-roll suppressed corrections.\footnote{A large body of work has been devoted to the study of conformal symmetry and inflation, see \cite{Creminelli:2011mw,Antoniadis:2011ib,Bzowski:2011ab,Maldacena:2011nz,Mata:2012bx,Arkani-Hamed:2015bza,Baumann:2015xxa} for a partial list. When appropriate, we will give the specific references in the body of the paper.} This is because, in this regime, the theory of inflaton fluctuations is approximately invariant under de Sitter isometries, which at late times, on superHubble scales, are isomorphic to the 3-dimensional Euclidean conformal group. In \refsec{s:softth}, we use these symmetries to fully fix the shape and the amplitude of the spectrum and bispectrum in the conformal limit.   
 For instance, consider the equal-time power spectrum of inflaton perturbations $ \varphi\equiv \phi-\bar\phi $ around a background $ \bar\phi $. On superHubble scales, the time dependence of $\varphi$ is the same as that of the background. Dilation symmetry fixes the $ k $-dependence in terms of the time dependence, yielding 
\be
\ex{\varphi_\k(\tau)\varphi_{-\k}(\tau)}'=C\frac{(k\tau)^{-\eta}}{k^{3}}\,,
\ee 
where $ C $ is a constant and a prime indicates that we dropped $ \left( 2\pi \right)^{3} $ and a Dirac delta function (we present our conventions below).
Notice that, for $ \eta \neq 0 $, $ \varphi $ evolves on superHubble scales, as expected. Using $ \varphi=-\zeta \sqrt{2\e} \Mpl$, we can convert to the curvature perturbations $ \zeta $, which are conserved on superHubble scales. We choose to convert to $ \zeta $ at some late conformal time $ \tau=\tau_{\ast} $, the \textit{same} for every wavenumber $ k $. We find
\bea
\ex{\zeta_{\k}\zeta_{-\k}}'=\frac{\tilde C \tau_{\ast}^{-\eta }} {k^{3+\eta} }\,,
\eea
where we have absorbed the $ k $-independent factor $ \e(\tau_{\ast}) $ into the constant $ \tilde C $. The spectral tilt is then easily seen to be $ \eta $, in agreement with \refeq{srns} for $ \e\ll\eta $. Notice that we did not have to solve the constraint equations of GR. Also, the $ \eta $ contribution to the observed spectral tilt $ n_{s} $, which is the largest one since $ \e\ll \eta $, does \textit{not} signal a breaking of dilation invariance during inflation. Rather, it is a precise consequence of the invariance of $ \varphi $ correlators under dilations (but not of $ \zeta $ correlators).

The bispectrum of $\varphi$ is also constrained by de Sitter isometries; it is the sum of two different shapes with arbitrary coefficients. To fix them we calculate the squeezed limit and impose that it matches the squeezed limit of the $ \varphi $ bispectrum derived using a background-wave argument. This is analogous to Maldacena's consistency condition in comoving gauge \cite{Maldacena:2002vr}.\footnote{One crucial difference is that our derivation is valid only in perturbation theory for the short modes, which is sufficient for our purposes. This is a weaker result than the standard $ \zeta $ soft theorem.} This uniquely fixes the three-point function of $\varphi$ (see \refeq{B32sym}). Performing the second order gauge transformation from $\varphi$ to curvature perturbations $\zeta$ gives 
\be\label{bisp2}
\langle \zeta_{\k_{1}} \zeta_{\k_{2}} \zeta_{\k_{3}}\rangle &=& \left( \frac{H^2}{4 \e\Mpl^2} \right)^{2}_{\ast} \Bigg\{\frac{\eta_{\ast}}{2} \frac{k_1^3 + k_2^3+k_3^3}{k_1^3 k_2^3k_3^3}+ \frac{\dot{\eta}_*}{2H_*}\frac{1}{k_1^3k_2^3k_3^3}\times \\
&&   \Big[(-1+\gamma_E+ \log{(-K \tau_*)})\sum_{i=1}^3 k_i^3-\sum_{i\neq j}k_i^2 k_j +  k_1 k_2 k_3\Big]\Bigg\}\,, \nonumber
\ee
where all time-dependent factors are evaluated at $\tau_*$ independently of $ k $. In \refsec{s:bisp}, we show that \refeq{bisp2} agrees with the direct calculation using the in-in formalism \cite{Burrage:2011hd}, and satisfies the squeezed limit consistency relation of Maldacena to next-to-leading order. An alternative derivation of the bispectrum using the wave functional formalism is presented in \refsec{sec:MMM}.

The non-Gaussian shape appearing in the second line of \refeq{bisp2} has been previously derived for spectator fields in de Sitter in \cite{Zaldarriaga:2003my,Seery:2008qj,Creminelli:2011mw,Kundu:2014gxa}. Our new result is to point out that these shapes with specific relative coefficients follow from symmetry arguments, and describe non-Gaussian features of primordial fluctuations in single-field inflation in the conformal limit. We recognize the first term in \refeq{bisp2} as the usual local shape with coefficient $ (n_{s}-1)/2\sim \eta/2 $, which is not locally observable to leading order in derivatives \cite{Pajer:2013ana}. The second term, proportional to $ \dot\eta $, is the next-to-leading order result, first derived in \cite{Burrage:2011hd}. We dub its $ k $-dependence the \textit{conformal shape}, since it is invariant under special conformal transformations (see \refsec{s:confinv}) and we show it in \reffig{shapes}.

In the conformal limit of inflation, the following simple relation arises between the amplitude of the conformal shape and the running of the power spectrum ($\alpha_s = d n_s/d\log k$):
\beq\label{equ:ncc}
\boxed{f^{\rm{conf.}}_{NL} =-\frac{25}{36} \alpha_{s}}\,.
\eeq
This is a new single-field consistency relation, which can in principle be tested within the scalar sector alone. Given the small expected size of the running, this relation will not be tested in the near future. Nevertheless, it is worth keeping in mind that the most natural, non-informative prior on $ r $ is a log-flat prior extending all the way to $ 10^{-55} $ or less. With this prior, the tensor consistency condition $r=-8n_t$ might well be even harder to test than this new, scalar one given by \refeq{equ:ncc}.\\

Finally, a few additional new results are scattered around the paper. For the convenience of the reader, we provide here an executive summary:

\begin{itemize}
\item We show in \refsec{s:bisp} how to derive the bispectrum in comoving gauge, without any field redefinition. This clarifies when and why the \textit{boundary terms} in the $ \zeta $ action are important. In particular, they need to be included to obtain constant correlators of $ \zeta $ on superHubble scales (see also \cite{Burrage:2011hd}).

\item In \refapp{a:A}, we note that the GR constraint equations are easier to solve in flat gauge. Since the conformal limit implies the decoupling limit (see \refeq{declimit}), we can use the hierarchy between the scalar and gravitational perturbations to solve the constraints in an $\Mpl\rightarrow \infty$ expansion, but \textit{to all orders in the field fluctuations}. The structure of the solutions to the constraint equations in comoving gauge is then a simple consequence of the change of coordinates from flat to comoving gauge.
\item It is well-known that the constraint equations can be solved to first order in perturbations if one is interested in the cubic action \cite{Maldacena:2002vr}. We prove in \refapp{constrtoorder} that the constraint solution to order $ n $ is sufficient to derive the action to order $ (2n+1) $, failing for the first time only at order $ (2n+2) $. This is a stronger result than the one proven in \cite{Chen:2006nt}, where the $ n $-th order constraint solution was proved to be sufficient for the action only up to order $ n+2 $. In particular, the known solution of the constraint equations to order 2 (see e.g. \cite{Arroja:2008ga}) can already be used to derive the action to 5th order.
\end{itemize}

{\bf Notation and conventions} We use natural units, $c=\hbar=1$, with reduced Planck mass $\Mpl^2\equiv1/8\pi G$. Our metric signature is $(-+++)$. Overdots and primes will denote derivatives with respect to physical time $t$ and conformal time $\tau$, respectively. The conformal time $\tau$ is defined by $dt \equiv a(\tau)d\tau$. 
We use the Hubble slow-roll parameters defined by
\be\label{defSR}
\e\equiv - \frac{\dot H}{H^2}\,,\quad \eta\equiv \frac{\dot\e}{H\e}\,,\quad \xi_{n\geq3}\equiv \frac{\partial \ln\xi_{n-1}}{\partial N}\,,
\ee
where $ \xi_{2}\equiv \eta $ and $ dN=Hdt $ is the number of e-foldings. We denote by the same symbol $N$ the lapse function in ADM decomposition, however the distinction between the two is always clear from the context. We use the label $ _{V} $ to indicate the potential slow-roll parameters, defined by
\be
\e_{V}\equiv \frac{\Mpl^{2}}{2}\left( \frac{V'}{V} \right)^{2}\,,\quad \eta_{V}\equiv\Mpl^{2}\frac{V''}{V}\,.
\ee
We indicate by $k$ the magnitude of the comoving wavenumber $\textbf{k}$. Our Fourier conventions are
\begin{equation}
F(\v{x})=\intk \tilde{F}(\v{k})e^{i\v{k}\cdot \v{x}}, \quad \text{where we use the shorthand} \quad \intk\equiv  \int \frac{d^3\v{k}}{(2\pi)^3}.
\end{equation}
A prime on a correlator indicates that we dropped the Dirac delta function and a factor of $(2\pi)^3$, 
\be
\ex{\varphi(\k_{1})\dots\varphi(\k_{n})}\equiv\left(  2\pi\right)^{3}\delta^{3}\left(  \sum \k_{i}\right)\,\ex{\varphi(\k_{1})\dots\varphi(\k_{n})}'\,.
\ee

 
\section{The Conformal Limit of Inflation}\label{s:2}

In this section, we define the conformal limit of inflation and summarize the simplifications that it implies for the solutions of the GR constraint equations. Technical details are collected in \refapp{a:A} and \refapp{PotinSRpar}.


\subsection{The Conformal Limit}\label{ss:conflim}

We are interested in the simplest model of single-field inflation: a canonical, minimally coupled scalar field $ \phi $ with an arbitrary potential $ V(\phi) $, as in the action \refeq{equ:infac}. According to observations, our universe is well described by the regime $ \e\ll \eta $, as discussed around \refeq{data}. 

The predictions of this model can be obtained by working at zeroth order in $ \e $ but keeping $ \eta $ and higher slow-roll parameters. More precisely, we start considering the de Sitter limit, namely $ \varepsilon\rightarrow 0 $. Since we want to keep the amplitude of primordial scalar perturbations $  H^{2}/\left(\varepsilon\Mpl^{2} \right) $ finite, we need to demand also $ H/\Mpl\rightarrow 0 $. Hence, we consider the limit
\bea\label{declimit}
\boxed{ \varepsilon,\frac{H}{\Mpl}\rightarrow 0 \quad \text{with}\quad \frac{H^{2}}{\varepsilon \Mpl^{2}}, \, \eta,\,\xi_{n}, \cdots ~\mathrm{finite}}\,.
\eea
One intuitive way to think about this limit is to keep $ H $ constant and send $ \Mpl $ to infinity. It is then clear that this is a \textit{decoupling limit}, in which the metric becomes non-dynamical (i.e., a classical background). We will see that it is consistent to work in this decoupling limit in flat gauge, and convert to curvature perturbations at the end of the calculation. As we discuss shortly in \refsec{ssec:staro}, \textit{this limit includes all viable small-field models of inflation ($ \Delta \phi\ll \Mpl $) as well as Starobinsky-like inflation}.

Since gravity becomes non-dynamical in the limit \refeq{declimit}, after choosing spatially flat gauge the action reduces to that of a scalar field in de Sitter. The theory for perturbations around an inflationary background $ \bar\phi(t) $ is 
\be
S=-\int d^{3}xdt\, e^{3Ht}\, \left[\frac{1}{2}  \partial_{\mu}\varphi\partial^{\mu}\varphi+ \sum_{m=2}^{\infty}\frac{\varphi^{m}}{m!}V^{(m)}(\bar\phi)\right]\,,
\ee
where $ V^{(m)} $ is the $ m $-th derivative of the potential with respect to $ \phi $. The time dependence of the background $ \bar\phi(t) $ induces a time dependence of the interaction coefficients $ V^{(m)}(\bar\phi) $ that breaks the boost and dilation isometries of de Sitter. On the other hand, this breaking of de Sitter isometries is suppressed by the slow roll parameters $\eta \sim \xi_{n} $ and can therefore be neglected to leading order. To see this, let us write $ V^{(m)} $ in terms of Hubble slow-roll parameters, as discussed in \refapp{PotinSRpar}. One finds that $ V^{(m)} $ is a polynomial in $ \eta $ and $ \xi_{n} $, with potentially negative half-integer powers of $ \e $. The time dependence of $ V^{(m)} $ is then given by the time dependence of the slow-roll parameters, which is slow-roll suppressed 
\be
\xi_{n}(N)&\approx&\xi_{n}(N_{\ast})\left[  1+\left.\frac{\partial_{N} \xi_{n}}{\xi_{n}}\right|_{N_{\ast}}\left(  N-N_{\ast}\right)\right]=\xi_{n}(N_{\ast})\left[  1+\O\left(  \xi_{n+1}\right)\right]\,,
\ee
hence proving our claim.

This suppression of the breaking of de Sitter isometries is unique to a canonical scalar field.\footnote{By canonical scalar field we mean \refeq{equ:infac}, but of course any theory of the form $ G(\phi)(\partial\phi)^{2} $ can always be brought into canonical form by a field redefinition.} If the scalar field were non-canonical, e.g. some $ P(X) $ model, the time dependent background would break de Sitter isometries by an amount that is \textit{not} suppressed by the slow-roll parameters. This is easy to see for example for the speed of sound $ c_{s}\neq 1 $. Inflaton perturbations $ \varphi $ propagate on the sound cone defined by $ c_{s} $, but this is not invariant under de Sitter boosts, which reduce to Minkowski boosts at short distances and leave only the light cone invariant.\footnote{For a brief discussion of the breaking of conformal symmetry induced by a nontrivial speed of sound, see App. A of \cite{Baumann:2015xxa}.} Therefore we need to assume that the scalar field is canonical, as in \refeq{equ:infac}. In particular, all coefficients in the Effective Field Theory of Inflation \cite{Cheung:2007st} that parameterize deviations from the vanilla slow-roll, canonical inflation are assumed to be negligible. Non-canonical models are further discussed in \refsec{disc}.

The assumptions of a canonical scalar field plus the limit \refeq{declimit} define the \textit{conformal limit of inflation}. In this limit, de Sitter isometries are unbroken (acting naturally in flat gauge) and all correlators of the inflaton perturbations (and of the graviton) must be de Sitter invariant.

For later reference, let us write the background equations of motion for the attractor FLRW solution, which will be quasi-de Sitter space in our case. They are
\bea\label{beom}
&&\dot{\bar\phi}^2=-2\Mpl^2\dot H\nonumber\, ,\\
&&V(\bar\phi)=\Mpl^2(3H^2+\dot H)\approx 3\Mpl^2H^2 ~~ \mathrm{and}\\
&&\ddot{\bar\phi}+3H\dot{\bar\phi}+V'(\bar\phi)=0\, . \nonumber
\eea
We see that in the limit of \refeq{declimit} the potential overwhelms the kinetic energy of the background field. Nonetheless, the kinetic energy is still finite, which is why the scalar fluctuations transform into curvature perturbations under the appropriate gauge transformation.

 
\subsection{Starobinsky Inflation as an Explicit Example}\label{ssec:staro}

The skeptical reader might wonder whether it is consistent to send $ \e $ to zero but keep $ \eta $ fixed, as we advocate in the conformal limit \refeq{declimit}. After all $ \eta\propto\dot\e $ and one might worry that $ \e $ soon becomes sizable. There are various ways to convince oneself that this is not an issue. First, consider the following explicit class of examples
\bea\label{scaling}
\e=\frac{\e_{0}}{\left(  -N\right)^{\beta}}\then \eta=\frac{\beta}{(-N)}\,,\quad \xi=\xi_{n\geq 3}=\frac{1}{ (-N)}\,,
\eea
where $ \beta>0 $, $N$ denotes the number of e-folds ($ dN=+Hdt $) going from around $ -N\sim60 $ at CMB scales to $ -N=1 $ at the end of inflation and $ \e_{0}\ll1 $ is a small parameter. For $ \beta>1 $ one indeed finds $ \varepsilon\ll \eta,\xi,\xi_{3}, \dots\ll 1$ at CMB scales $ -N\sim 60 $. It is therefore perfectly natural for a general model of inflation to have a hierarchy between the first slow-roll parameter $ \e $ and all the others. In fact, if we furthermore impose that the spectral tilt takes the measured value, we select $ \beta \simeq 2 $ (with small corrections due to the uncertainty in the duration of inflation). The examples in \refeq{scaling} actually describe two very popular classes of models: Starobinsky-like inflation~\cite{Starobinsky:1980te} (and its modifications, e.g. $ \alpha $ attractors \cite{Kallosh:2013tua}) and small-field models. Consider, for example, the Starobinsky-like potential
\be\label{staro}
V=V_0 \left( 1-e^{-\phi/M} \right)^2\,,
\ee
where $M<M_{\rm Pl}$ is some mass scale. The slow-roll parameters in this model are precisely as in \refeq{scaling} with $ \e_{0}=2(M/\Mpl)^{2} $ and $ \beta=2 $. Therefore, for $M\ll M_{\rm Pl}$, one finds $ \e\ll\eta $, parametrically. For small-field models, with subPlanckian field displacement, we cannot provide a general formula, but the estimate using the Lyth bound in \refeq{smallfield} also leads to the same scaling as in \refeq{scaling} with $ \e_{0}\sim (\Delta \phi/\Mpl)^{2}\ll1 $. Notice that, in the cases discussed above, one finds the stronger condition $ \e\ll\eta^{2} $. This implies that $ \e $ can be neglected even when including the first subleading order correction in $ \eta $, namely the order $ \eta^{2} $. Moreover, one can always choose parameters (in a natural way) such that $ \e\ll \eta^{p} $ for $ p\gtrsim 2 $.

Alternatively, let us just Taylor expand $ \e $ around some arbitrary $ N_{\ast} $
\be
\e(N)-\e(N_{\ast})&=&  \left.\frac{\partial \e}{\partial N}\right|_{N_{\ast}} \left(  N-N_{\ast}\right)+\left.\frac{\partial^{2} \e}{\partial N^{2}}\right|_{N_{\ast}}\frac{\left(  N-N_{\ast}\right)^{2}}{2} +\mathcal{O}\left(  \partial_{N}^{3}\e \right)\nonumber\\
&=& \e \left[ \eta \left(  N-N_{\ast}\right)+ \eta \xi_{3}\frac{\left(  N-N_{\ast}\right)^{2}}{2} +\mathcal{O}\left(\eta^{3}, \eta^{2}\xi_{3},\eta \xi_{3}\xi_{4}, \e \right) \right]\,,
\ee
where all the slow-roll parameters are evaluated at $ N_{\ast} $. In the second line we see that the evolution of $ \e $ is itself suppressed by $ \e $. So, as long as $ \e(N_{\ast})\rightarrow 0 $, it is consistent to neglect it. 

Summarizing, one should keep in mind that \textit{Starobinsky inflation and its modifications, as well as all viable small-field models are specific cases of the conformal limit} we study in this paper, satisfying the stronger condition  $ \e\ll\eta^{2} $. This ensures that the subleading correction to the bispectrum \eqref{bisp2} is still large compared to deviations from conformal symmetry, expected at order $ \e $.

 
\subsection{Constraint Equations in the Conformal Limit}

In this subsection, we discuss some interesting results about the constraint equations in the conformal limit. The derivations and a detailed discussion are collected in \refapp{a:A}. 

In the conformal limit, the constraint equations simplify considerably. Written in flat gauge, they read 
\bea
&&\Mpl^2\left(R^{(3)}-6H^2-N^{-2}h^{ij}h^{kl}(E_{ik}E_{jl}-E_{ij}E_{kl})\right)-\nonumber\\&&-\Mpl^0\left(N^{-2}(\dot{\bar\phi}+\dot\varphi-N^i\partial_i\varphi)^2+(-\dot{\bar\phi}^2+2V'(\bar\phi)\varphi+\cdots )+h^{ij}\partial_i\varphi\partial_j\varphi \right)=0 \, ,\\
&&\Mpl^2\left(\nabla_a\left(N^{-1}(h^{ab}E_{bi}-\delta^a_i h^{bc}E_{bc})\right)\right)+\Mpl^0\left( N^{-1}\partial_i\varphi(N^j\partial_j\varphi-\dot{\bar\phi}-\dot\varphi)\right)=0\, .
\eea

An obvious observation is that, in the $\Mpl^2\to\infty$ limit, the constraint equations admit the trivial solutions 
\bea\label{tsc}
N=1 +O(\Mpl^{-2}), ~~N_i=O(\Mpl^{-2}) \, .
\eea
In fact, we can use $\Mpl^{-2}$ as a small expansion parameter, rather than the field perturbations $\varphi$, as is usually the case. This allows us to solve the constraint equations to all orders in $\varphi$ to subleading order in $\Mpl^{-2}$. We present the solution in \refapp{a:A}.

If we write the constraint equations in comoving gauge, $\Mpl^2$ cancels out, so there is no perturbation theory in the inverse Planck mass. The decoupling limit corresponds to the de Sitter limit, so if we take $\varepsilon \to 0$, one would be tempted to guess that the constraint equations have solutions as simple as \refeq{tsc}. This is not the case. Even in the $\varepsilon \to 0$ limit, the constraint equations have rather non-trivial solutions in comoving gauge. Nonetheless, we can obtain them by changing coordinates from flat gauge (with the background FLRW metric, due to \refeq{tsc}) to comoving gauge. Finding the coordinate transformation from $\varphi$ to $\zeta$ gauge is much easier than tackling the constraint equations, so this is a more economical route to solving the constraint equations. We checked explicitly that, in the conformal limit, to second order in $\zeta$, the solutions to the constraint equations are given by a change of coordinates from the flat FLRW metric to the comoving coordinates. 

Finally, we point out (see \refapp{constrtoorder}) that solving the constraint equations up to a certain order in perturbation theory goes a long way in finding the perturbative action. Namely, if we solve the constraint equation to order $n$ in perturbations, that is enough to determine the perturbative action to order $2n+1$. This result is stronger than the one derived in \cite{Chen:2006nt}, where it was proven that the $ n $-th order solution of the constraints is sufficient for the action at order $ (n+2) $. 


\section{The Spectra from Symmetries}\label{s:softth}
As discussed in \refsec{s:2}, in the conformal limit \refeq{declimit}, inflaton correlators are invariant under de Sitter isometries at leading order in slow-roll. Six of them are manifest. Invariance under spatial translations and spatial rotations tells us that the three-point function is proportional to a momentum conserving Dirac delta function and that it only depends on scalar products of the momenta. The consequences of the additional four isometries are less obvious, so we review them here. If we write the de Sitter line element in flat slicing,
\beq
\dd s^2=\frac{-\dd\tau^2+\dd \x^2}{(H\tau)^2} \, ,
\eeq
it is easy to verify that these isometries are given by (in infinitesimal form)
\begin{align}
\label{eq:dSdil}
& \qquad \tau \,\to\, \tau \hskip 1pt (1+ \lambda) \ , \quad \hspace{1 cm}  \x \,\to\, \x \hskip 1pt (1+\lambda)\ , \\
\label{eq:dSsct}
& \qquad \tau \,\to\, \tau(1 - 2\hskip 1pt \b\hskip 1pt.\hskip 1pt \x\hskip 1pt)\ , \qquad \x \,\to\, \x-2(\hskip 1pt \b\hskip 1pt.\hskip 1pt \x\hskip 1pt) \hskip 2pt \x+(\x\hskip 1pt{}^2-\tau^2)\hskip 1pt \b\ ,
\end{align}
where $\lambda$ is a real infinitesimal parameter and $\b$ an infinitesimal, 3-dimensional vector. The consequences of these symmetries for scalar and tensor correlators have already been studied in the literature \cite{Larsen:2003pf,Antoniadis:2011ib,Maldacena:2011nz,Bzowski:2011ab,Creminelli:2011mw,Mata:2012bx,Kehagias:2012pd,Kundu:2014gxa,Bzowski:2013sza,Kundu:2015xta,Arkani-Hamed:2015bza}, and we will use some of these results. 

In practice we are always interested in the late-time correlators of inflaton perturbations $ \varphi(\k,\tau)$, when all the modes are outside the horizon. In the limit $ -\tau\ll 1/k \sim x $ we can neglect the $\tau^2$ term in \refeq{eq:dSsct}, and transformations of $\x$ are isomorphic to the infinitesimal generators of dilations and special conformal transformations of the conformal group acting on the spatial slice. For massive fields the transformation of $\tau$ can be taken into account by assigning a scaling dimension $\Delta$ to the operator $\varphi$; it is fixed by looking at the late time dependence $\varphi(\k,\tau) \sim \tau^\Delta \varphi(\k)$. Therefore, in the late time limit, dS invariant correlators must have the same form as correlators of some CFT where fields have conformal weight $\Delta$ fixed by the inflaton mass. This imposes strong constraints on the power spectrum and the bispectrum. In this section we show how their shape is entirely fixed by this symmetry. 

 
\subsection{Power Spectrum}\label{ss:ps}

Let us first focus on the power spectrum and re-derive the spectral tilt $n_{s} $ by converting inflaton perturbations $ \varphi $ into curvature perturbations $ \zeta $. We can treat the inflaton as a free scalar field in de Sitter with mass $V''$. At late times de Sitter isometries impose that the power spectrum has the following form \cite{Arkani-Hamed:2015bza} 
\be\label{timedepphi}
\ex{\varphi_\k(\tau)\varphi_{-\k}(\tau)}'&=& \frac{H^{2}}{k^{3}}\frac{1}{\pi} \left[  \Gamma\left(  \frac{3}{2}-\Delta_-\right)^2\left(-k\tau\right)^{2\Delta_-}+\left(\Delta_-\to\Delta_+\right)\right]\,,
\ee
where $ \Delta_{\pm}\equiv\frac{3}{2}\pm \sqrt{9/4 - V''/H^2} $. We are interested in the limit where the inflaton mass is small and we want to keep only the growing mode. At leading order in slow-roll $\Delta_-$ is given by
\bea\label{appro}
\Delta_{-}= \frac{V''}{3H^{2}}+\O\left( \frac{ (V'')^{2}}{H^{4}}\right) \simeq  \eta_{V}\simeq -\frac{\eta}{2} \,.
\eea
Using $\Gamma(3/2)^2=\pi/4$, the power spectrum simplifies to
\be
\ex{\varphi_\k(\tau)\varphi_{-\k}(\tau)}'&=& \frac{H^{2}}{4k^{3}} \left(-k\tau\right)^{-\eta} \,.
\ee

We now need to make a gauge transformation from $\varphi$ to $ \zeta $, which is conserved on superHubble scales. This is achieved by $ \zeta=-\varphi/\Mpl\sqrt{2 \e} $ where the slow-roll parameter $ \e\equiv - \dot H/ H^{2} $ is evaluated at a chosen time $ \tau $. We make two different choices and show that they give the same result. First, we can do the conversion at the Hubble crossing of each mode, namely at $ \tau=\tau_{\text{H.c.}}\equiv -1/k $. The result is the usual expression
\be\label{uno}
\ex{\zeta_{\k}\zeta_{-\k}}'=\frac{H^{2}}{4 \e \Mpl^{2}}\Bigg|_{\text{H.c.}}\frac{1}{k^{3}}\,,
\ee
where ``H.c.'' indicates that the time-dependent quantities $ H $ and $ \e $ should be evaluated at Hubble crossing. This procedure has the advantage that one does not need to know the slow-roll suppressed time dependence of  $ \varphi $ perturbations on superHubble scales, since the term $ (-k\tau)^{2\Delta_{-}} $ (see \refeq{timedepphi}) becomes unity. On the other hand, $ \tau $ is now $ k $-dependent and therefore the spectral tilt of the power spectrum hides inside the time dependence of $ H $ and $ \e $. This is a disadvantage since the spectral tilt remains implicit. 

In our discussion, de Sitter isometries have already fixed the $\varphi$ time dependence to be $ (-k\tau)^{-\eta} $. Therefore, we might as well proceed in a different but equivalent way. We evolve each mode until some late fixed time $ \tau_{\ast} $, the same for every mode and such that $ -k\tau_{\ast}\ll 1 $ for every $ k $ of interest. Using $ \varphi=-\zeta \sqrt{2\e_*}\Mpl $ we find
\bea\label{due}
\ex{\zeta_{\k}\zeta_{-\k}}'=\frac{H^{2}}{4 \e \Mpl^{2}}\Bigg|_{\ast}\frac{1}{k^{3+\eta}}\tau_{\ast}^{-\eta}\,,
\eea
where, as in \refeq{uno}, this is the asymptotic, time-independent value of the $ \zeta $ power spectrum. This formulation has the advantage of making the tilt explicit, since now, the factor $ H^{2}/\e $ is a constant that does not depend on $ k $. The tilt in \refeq{due} agrees with the standard result $ 1-n_{s}=2\e+\eta $ in the conformal limit, for $ \e\rightarrow 0 $.

  
\subsection{Conformally Covariant Shapes of the Bispectrum}\label{s:confinv}

Let us turn to the discussion of how de Sitter isometries, to which we also refer as \textit{conformal symmetry}, constrain the bispectrum of $ \varphi $. We find that the bispectrum is fixed up to two multiplicative constants. As we then show in \refss{ssec:together}, these constants can be fixed using the consistency condition for the squeezed limit of the three-point function which we derive in the following section. 

The implications of the covariance under de Sitter isometries in momentum space can be derived from transformations in Eqs.~(\ref{eq:dSdil})--(\ref{eq:dSsct}). For the three-point correlation function of $\varphi$ they simplify to~\cite{Maldacena:2011nz,Bzowski:2013sza}
\be 
\left[-3(\Delta-2)+\sum_{a=1}^3 k_a  \frac{\partial}{\partial k_a}\right]\langle \varphi(\textbf{k}_1)\varphi(\textbf{k}_2)\varphi(\textbf{k}_3)\rangle'\overset{!}{=}&{\rm``local"}\,,
\label{D3pt}\\ 
\sum_{a=1}^3\left[\b\cdot\k_a \left(\frac{\partial^2}{\partial k_a^2}- \frac{2(\Delta -2)}{k_a}\frac{\partial }{\partial k_a} \right) \right] \langle \varphi(\textbf{k}_1)\varphi(\textbf{k}_2)\varphi(\textbf{k}_3)\rangle'\overset{!}{=}&0\,.
\label{SCT3pt1}
\ee
We keep the dependence in $\Delta$ explicit for now, but for inflation we will eventually be interested in $\Delta=0$.  \refeq{D3pt} constraints the overall momentum scaling of the three-point function to
\beq
\langle \varphi(\textbf{k}_1)\varphi(\textbf{k}_2)\varphi(\textbf{k}_3)\rangle' \overset{!}{=} k_1^{3(\Delta -2)}F\left(\frac{k_2}{k_1},\frac{k_3}{k_1}, \log(k_1+k_2+k_3) \right),
\eeq
with $F\left(\frac{k_2}{k_1},\frac{k_3}{k_1}, \log(k_1+k_2+k_3) \right)$ an arbitrary function with no overall momentum scaling. The term ``local" on the right hand side of the dilation equation requires some explanation. Introducing a logarithm in $F$ implies introducing an arbitrary scale in the correlator. This follows \cite{Maldacena:2011nz, Coriano:2013jba} from the assumption that we are always allowed to integrate by parts in performing Fourier transforms, which is necessary to substitute $\partial_\x\to i \k$. If the Fourier transform is marginally convergent, we might miss a boundary term, which produces an anomalous (local) violation of dilation invariance. A useful example to keep in mind is that of the two point function of the stress tensor in a CFT in $d$ even dimensions. The dilation constraint would imply that $\langle T_{ij}(\k) T_{ij}(-\k) \rangle' \sim k^{d}$, which is a contact term. The correct two point function is given by $\langle T_{ij}(\k) T_{ij}(-\k) \rangle' \sim k^{d} \log k$. Now, the action of the dilation operator will produce a $k^d$, which is a contact term. In summary, dilation should fix the overall scaling of the correlator up to logarithms. The action of the dilation operator on the three-point function produces a term that corresponds to a contact term in the putative dual CFT. In the case at hand, this is the local non-Gaussian shape \refeq{local}.

To constrain the three-point function using the special conformal isometry, let us use our freedom in picking $\b$ to write \refeq{SCT3pt1} in a more convenient form. If we choose $\b$ such that $\b\cdot \k_3=0$, then $\b\cdot\k_1=-\b\cdot\k_2$.
This simplifies \refeq{SCT3pt1} considerably, and we obtain
\beq
 \left[K_a - K_b \right]\langle \varphi(\textbf{k}_1)\varphi(\textbf{k}_2)\varphi(\textbf{k}_3)\rangle'\overset{!}{=}0\,.
\label{SCT3pt2}
\eeq
with $a\neq b $ and 
\beq
K_a \equiv \left[\frac{\partial^2}{\partial k_a^2}-\frac{2(\Delta-2)}{k_a}\frac{\partial}{\partial k_a} \right].
\label{SCT3pt3}
\eeq
The solutions to both \refeq{D3pt} and \refeq{SCT3pt2}, up to a multiplicative constant, are given by \cite{Bzowski:2013sza} 
\be\label{tripleK}
\langle \varphi(\textbf{k}_1)\varphi(\textbf{k}_2)\varphi(\textbf{k}_3)\rangle' &= & (k_1k_2k_3)^{\Delta -3/2}\\
&&\quad \times \int_{0}^{\infty} dx~x^{3/2-1}\text{K}_{\Delta -\frac{3}{2}}(k_1 x)\text{K}_{\Delta -\frac{3}{2}}(k_2 x)\text{K}_{\Delta -\frac{3}{2}}(k_3 x)\,. \nonumber
\ee
This solution is known as the triple K integral. Only modified Bessel functions with half-integer $\Delta-\frac{3}{2}$ give us expressions for $\text{K}_{\Delta-\frac{3}{2}}(k_ix)$ in terms of elementary functions. This limits the number of exact calculations we can perform.

The integral in \refeq{tripleK} might not converge and some regularization scheme is necessary. When all variables are real, \refeq{tripleK} converges for
\begin{equation}
\frac{3}{2}>3 \left|\Delta - \frac{3}{2} \right|+2\,.
\end{equation}
To construct the three-point correlation function from symmetries for $m^2=0$, we employ the following regularisation scheme \cite{Bzowski:2013sza}:
\begin{equation}
\begin{split}
3\rightarrow 3+2\delta,&~~~~~ \Delta \rightarrow \Delta + \delta\,.
\end{split}
\label{regularization}
\end{equation}
Note that $\delta$ is just an expansion parameter, as is done in dimensional regularization\footnote{We deviate from the standard notation and choose $\delta$ to avoid confusion with the slow-roll parameter $ \e $.}. For the massless case in 3 (spatial) dimensions, we have $\Delta=0+\O(\eta)$. The modified Bessel functions then simplify to
\begin{equation}
K_{3/2}(x) = K_{-3/2}(x)=\sqrt{\frac{\pi}{2}}(1+x)\frac{e^{-x}}{x^{\frac{3}{2}}}\,.
\label{K32}
\end{equation}
Using the regularisation scheme in \refeq{regularization}, we obtain
\be
&\lim_{\delta\rightarrow0}\displaystyle\int_0^{\infty}dx~x^{\frac{3}{2}-1+\delta}K_{-3/2}(k_1 x)K_{-3/2}(k_2 x)K_{-3/2}(k_3 x)=\\
&=\lim_{\delta\rightarrow0}\left[ \left(\frac{\pi}{2}\right)^{\frac{3}{2}} \frac{(k_1+k_2+k_3)^{-\delta}}{(k_1 k_2 k_3)^{\frac{3}{2}}}(-2+\delta)\left(\sum_{i=1}^3 k_i^3 + \delta \left(\sum_{i\neq j}^3 k_i^2 k_j -k_1 k_2 k_3\right)+ \delta^2 k_1 k_2 k_3)\right)\Gamma(-3+\delta)\right]\,.\nonumber
\label{tripleK32}
\ee
At leading order in $\delta$, we obtain
\begin{equation}\label{local}
\langle \varphi(\textbf{k}_1)\varphi(\textbf{k}_2)\varphi(\textbf{k}_3)\rangle' =C_{\text{loc}} \frac{k_1^3+k_2^3+k_3^3}{k_1^3k_2^3k_3^3}\,,
\end{equation}
where, for future convenience, we introduced an arbitrary constant $ C_{\text{loc}} $ which is not fixed by conformal symmetry. \refeq{local} is the well-known local shape of the bispectrum. At the next to leading order in $\delta$, we obtain
\beq
\langle \varphi(\textbf{k}_1)\varphi(\textbf{k}_2)\varphi(\textbf{k}_3)\rangle' =C_{\text{con}}\, \frac{ \log{(K/k_{\ast})}\sum_{i=1}^3 k_i^3-\sum_{i\neq j}k_i^2 k_j +  k_1 k_2 k_3}{k_1^3k_2^3k_3^3}\,,
\label{equilog}
\eeq
where $ K\equiv k_{1}+k_{2}+k_{3} $. We dub \refeq{equilog} the \textit{conformal shape} of the bispectrum. To keep the argument of the log dimensionless, we subtracted a purely local term with coefficient $\ln(k_{\ast})$, which we can always do by redefining the coefficient of the local term in \refeq{local}. In the next section, we will find that $ k_{\ast} $ can be interpreted as the Hubble scale at some late time during inflation, $ k_{\ast}=-1/\tau_{\ast} $ in \refeq{B32}. It is straightforward to use \refeq{SCT3pt2} to check that both the local and the conformal shapes are separately invariant under special conformal transformations. In particular, note that neither $ \ln(K) $ times the local shape nor the other two terms in the conformal shape \refeq{equilog} are separately invariant. It is only the sum of the three terms that is invariant under special conformal transformations. 

Let us note that for $\Delta=0$ the invariance under dilation, \refeq{D3pt}, of the local shape is manifest, while that of the conformal shape is spoiled by the presence of the logarithmic term. This looks like a paradox and requires some explanation. Under the action of the dilation operator, the conformal shape produces precisely the local shape. Like in the discussion after \refeq{D3pt}, we have an anomaly due to neglecting the boundary contribution to the Fourier transform of the conformal shape. This is particularly clear if we interpret these three-point functions as coming from a putative CFT. To obtain the CFT three-point function from our shape functions, we strip the $(k_1 k_2 k_3)^3$ factors in the denominators of both of them. The conformal shape will give a CFT three-point function, and the action of dilations on this three-point function will give a sum of terms that only has support when two points coincide; in other words, dilations are not violated when all three points are separated. The local shape gives a CFT three-point function that has zero support when all three points are separated. In position space, it corresponds to terms of the form $\delta(\x-{\bf y})f(\x-{\bf z})+ {\rm perm.}$, etc.

The other way to see that there is no paradox is to restore the $\tau$ dependence in the logarithm and note that, for $\Delta=0$, the de Sitter isometries corresponding to dilations act on the equal-time correlation functions as
\be
\delta_\lambda \langle \varphi(\textbf{k}_1)\ldots \varphi(\textbf{k}_n)\rangle' = \lambda \left[ \tau \partial_{\tau} - 3(n-1) - \sum_{a=1}^n k_{ai} \partial_{k_{ai}} \right] \langle \varphi(\textbf{k}_1)\ldots \varphi(\textbf{k}_n)\rangle' =0\;.
\ee
This operator follows from \refeq{eq:dSdil} and it takes into account the fact that the rescaling of coordinates must be accompanied by a shift in the conformal time. As the $\tau$ dependence of the bispectrum is only appearing explicitly in the $\log(K\tau)\times{\rm local}$ term, we see that the dilation anomaly precisely cancels with the action of the $\tau \partial_\tau$ term. This is exactly what we expect, because the three point-function is de Sitter invariant. 

Before proceeding, we mention that a similar calculation can be done for the case $ \Delta=1 $, corresponding to $ V''=2H^{2} $. Using again the regularization scheme in \refeq{regularization}, we obtain
\be
&\lim_{\delta\rightarrow0}\displaystyle\int_0^{\infty}dx~x^{-1+\delta}K_{-1/2}(k_1 x)K_{-1/2}(k_2 x)K_{-1/2}(k_3 x)=\\
&=\lim_{\delta\rightarrow0}\left[ \left(\frac{\pi}{2}\right)^{\frac{3}{2}}\frac{(k_1+k_2+k_3)^{-\delta }}{\sqrt{k_1 k_2 k_3}}\Gamma(\delta)\right]\,.\nonumber
\label{tripleK12}
\ee
At leading order in $\delta$, we find
\be
\langle  \varphi(\textbf{k}_1)\varphi(\textbf{k}_2)\varphi(\textbf{k}_3)\rangle'=\frac{C_{1}}{k_1 k_2 k_3}\,.
\label{comp}
\ee
Similar to the $\nu = 3/2$ case, we can also expand \refeq{tripleK12} to next to leading order, where one obtains
\beq
\langle  \varphi(\textbf{k}_1)\varphi(\textbf{k}_2)\varphi(\textbf{k}_3)\rangle' = C_2 \frac{\log(K/k_*)}{k_1 k_2 k_3}\,.
\eeq
It is interesting to note that while both of these shapes are allowed by symmetry, in the explicit in-in calculations for massive fields with $\Delta=1$ it turns out that $C_2=0$. 


\subsection{A Soft Inflaton Consistency Relation}\label{ssec:soft}
So far we were able to show how symmetries fix the bispectrum to be a sum of two different shapes with amplitudes $C_{\rm loc}$ and $C_{\rm con}$. The next step is to fix these constants. This is possible due to a consistency condition that relates the squeezed limit bispectrum and the power spectrum. This condition is well known for $\zeta$ correlators. Here we re-derive it for the $\varphi$ bispectrum following the standard approach \cite{Creminelli:2004yq}.

When taking the squeezed limit of a $\varphi$ correlation function, i.e. taking one of the external momenta to be very small, one of the modes has a much smaller wavenumber than the others. This mode leaves the Hubble radius much earlier. To leading order in derivatives, the short modes perceive it as a change in the background\footnote{One may wonder whether this background fluctuation can change the unperturbed Hubble parameter and its time derivative. To see that this is not the case, let us solve perturbatively the background equations of motion \refeq{beom} to find $\delta H$ and $\delta \dot H$ as functions of $\varphi$. To linear order, we obtain
\bea
&&V'(\bar\phi)\varphi=\Mpl^2(6H\delta H+\delta \dot H)~~\mathrm{and} ~~2\dot{\bar\phi}\dot\varphi=-2\Mpl^2\delta \dot H \;.
\eea
The solution is
\bea
&&\delta H = \frac{1}{6\Mpl^2 H}\left[(-\ddot{\bar\phi}-3H\dot{\bar\phi})\varphi+\dot{\bar\phi}\dot\varphi\right]\, ,  \quad \mathrm{and} \quad\delta \dot H =- \frac{\dot{\bar\phi} \dot\varphi}{\Mpl^2}\, ,
\eea
where we used \refeq{beom} to write $V'(\phi)$ in terms of time derivatives of $\bar\phi$. These corrections vanish in the conformal limit \refeq{declimit} since the numerators are all finite, while the denominator is proportional to $\Mpl^2$.}. In this way, we can rewrite the squeezed limit of a correlation function as a field shift
\begin{equation}
\langle  \varphi_l(\textbf{x},\tau) \varphi_s(\textbf{x}_1,\tau) ...  \varphi_s(\textbf{x}_n,\tau) \rangle =  \langle \varphi_{l}(\v{x})\langle \tilde{\varphi}_s(\textbf{x}_1,\tau) ...  \tilde{\varphi}_s(\textbf{x}_n,\tau) \rangle_{\varphi_l}  \rangle,
\label{squeezed1}
\end{equation}
where the subscript $s$ and $l$ refer to small and long wavelength modes, respectively. Note that on the LHS of \refeq{squeezed1} we have a ($n+1$)-point function and on the RHS we have a $n$-point function. The last term in \refeq{squeezed1} is evaluated in the background of a $\varphi_l(\textbf{x}_{0})$ mode. We can Taylor expand the right hand side of \refeq{squeezed1}
\begin{equation}
\begin{split}
\left.\langle \tilde{\varphi}(\textbf{x}_1,\tau) ...  \tilde{\varphi}(\textbf{x}_n,\tau) \rangle\right|_{\varphi_l} =& \langle \varphi_s(\textbf{x}_1,\tau) ...  \varphi_s(\textbf{x}_n,\tau) \rangle\\
&+ \varphi_l(\textbf{x})\left[ \frac{\delta}{\delta \varphi_l(\textbf{x})} \langle \varphi_s(\textbf{x}_1,\tau) ...  \varphi_s(\textbf{x}_n,\tau) \rangle\right]_{\varphi_l = 0}+ \mathcal{O}(\varphi_l^2).
\label{squeezed2}
\end{split}
\end{equation}
The second term gives the leading non-vanishing contribution to \refeq{squeezed1}, which becomes
\begin{equation}
\begin{split}
\langle \varphi_l(\textbf{x}_1,\tau) \varphi_s(\textbf{x}_2,\tau)\varphi_s(\textbf{x}_3,\tau)\rangle &=\langle \varphi_l(\textbf{x}_1,\tau)\varphi_l(\textbf{z},\tau)\rangle\left[\frac{\delta}{\delta \varphi_l(\textbf{z})}\langle \varphi_s(\textbf{x}_2,\tau)\varphi_s(\textbf{x}_3,\tau)\rangle_{\varphi_l}\right]_{\varphi_l= 0} \,,
\label{squeezed4}
\end{split}
\end{equation}
where $\textbf{z} = (\x_2+\x_3)/2$. In Fourier space this becomes
\begin{equation}
\langle \varphi(\textbf{k}_1,\tau)\varphi(\textbf{k}_2,\tau)\varphi(\textbf{k}_3,\tau) \rangle'_{k_1 \ll k_2, k_3} \approx \langle \varphi_l(\textbf{k}_1,\tau) \varphi_l(-\textbf{k}_1,\tau)\rangle' \frac{\delta}{\delta \varphi_l}\left.\langle \varphi_s(\textbf{k}_2,\tau) \varphi_s(\textbf{k}_3,\tau)\rangle'\right|_{\varphi_l= 0}\,.
\label{squeezed5}
\end{equation}

To explore this relation in the conformal limit, we need to compute how a shift in the background by $ \varphi_{l} $ affects the power spectrum. The leading interaction in the slow-roll expansion is given by the cubic term in the potential, which couples long and short modes. The linear order effect of a long $ \varphi_{l} $ mode is to generate an effective mass term for $\varphi_s$
\be
\meff^{2}\equiv V'''(\bar{\phi})\varphi_l(\tau)\,,
\label{tdm}
\ee
as can be seen from
\beq
-\frac{V'''}{6}\varphi^3 \rightarrow -\frac{V'''}{6} \left(\varphi_s^3+3\varphi_l \varphi_s^2 +  3\varphi_l^2 \varphi_s +\varphi_l^3\right).
\eeq
Since $ \varphi_{l} $ evolves on superHubble scales, the effective mass of $ \varphi_{s} $ is time dependent. The general power spectrum for a field with a time dependent mass is hard to calculate, since we cannot solve the equation of motion for $ \varphi_{s} $. On the other hand, we are interested only in the \textit{linear} effect of $ \meff^{2} $, so we can treat $ \meff(t)^{2} $ perturbatively. Since we know the solution of the equations of motion for a scalar field in de Sitter, we know the Green's function of the equation of motion for $ \varphi_{s} $ when $ \meff=0 $. The linear effect of $ \meff^{2} $ is then obtained from an integral of the Green's function, which can be done analytically. We collect all the details of the calculation in \refapp{a:gfct} and only quote the final result here. The squeezed bispectrum one obtains by substituting the zeroth and first order mode functions \refeq{modefuncnu} and \refeq{parsol32} into \refeq{squeezed5} is
\beq
\langle \varphi(\textbf{k}_s,\tau)\varphi(\textbf{k}_s,\tau)\varphi(\textbf{k}_l,\tau)  \rangle' = \frac{H^2V'''}{6}\frac{\left[ -2 + \gamma_E + \log{(-2 k_s \tau)} \right]}{k_s^3 k_l^3}+\O\left(  \frac{k_{l}^{2}}{k_{s}^{2}}\right)\,.
\label{sB32}
\eeq

As mentioned before, the consistency relation \refeq{squeezed4} is valid for any value of the mass $ V'' $, but for a generic $ V'' $ we are not able to compute the derivative analytically, since we cannot perform the integral that gives the linear order $ \varphi_{l} $ correction to the power spectrum of $ \varphi_{s} $. Besides the case $ V''=0 $, which we just discussed, there are two more cases we are able to treat analytically: the case of a small mass, $ |V''|\ll H^{2} $, which is relevant for inflation, and the case of $ V''=2H^{2} $, which is not. For the latter, the squeezed bispectrum one obtains by substituting the zeroth and first order mode functions \refeq{modefuncnu} and \refeq{parsol12} into \refeq{squeezed5} is
\beq
\langle \varphi(\textbf{k}_s,\tau)\varphi(\textbf{k}_s,\tau)\varphi(\textbf{k}_l,\tau)  \rangle' = \frac{\pi H^2 V'''(\bar{\phi})  }{8}\frac{\tau^3}{k_s^2 k_l}+\O\left(  \frac{k_{l}^{2}}{k_{s}^{2}}\right)\,.
\label{sB12}
\eeq
As we will see, both \refeq{sB32} and \refeq{sB12} correctly reproduce the squeezed limit of the $ \varphi $ bispectrum computed with the in-in formalism \cite{Burrage:2011hd}, \refeq{B32} and \refeq{B12}. 

In the case of a small mass, $ V''\ll H^{2} $, discussed in \refapp{ssec:intdim2}, we find \refeq{lightmassB} for the squeezed limit bispectrum. This expression is rather involved and describes a subleading slow-roll correction, suppressed by $ \eta $, to what is already a small non-Gaussianity, so we do not discuss it further here.


\subsection{The Primordial Bispectrum}\label{ssec:together}

We have now all the ingredients to fix the full shape of the bispectrum. By taking the squeezed limit of \refeq{local} and \refeq{equilog} and comparing it with \refeq{sB32} we find that 
\be\label{Cs}
C_{\text{loc}}&=& \frac{H^2V'''}{12}  \left( -1 + \gamma_E   \right)\,,\quad C_{\text{con}}= \frac{H^2V'''}{12} \,.
\ee
Note that the presence of the logarithmic term in the conformal shape \refeq{equilog} is what allows us to compute two constants from a single squeezed limit. Had the logarithmic term not been there, only the sum of $C_{\text{loc}}  $ and $ C_{\text{con}} $ would have been fixed, but not their difference. We therefore find the \text{full}, equal-time bispectrum of $ \varphi $ from symmetries to be
\be\label{B32sym}
\langle  \varphi(\textbf{k}_1,\tau)\varphi(\textbf{k}_2,\tau)\varphi(\textbf{k}_3,\tau)\rangle' &=&\frac{H^2}{k_1^3k_2^3k_3^3}\frac{ V'''(\bar{\phi})}{12}\\
&& \hspace{-1cm} \times\left[(-1+\gamma_E+ \log{(-K \tau)})\sum_{i=1}^3 k_i^3-\sum_{i\neq j}k_i^2 k_j +  k_1 k_2 k_3\right]\,,\nonumber
\ee
where we made the time dependence $ \varphi(\k, \tau) $ explicit to emphasize that this bispectrum is time-dependent. This is expected since, even for $ V''=0 $, the cubic interaction $ V''' $ makes the inflaton perturbations $ \varphi $ evolve on superHubble scales. 

We can also fix the normalization of the massive case. Comparing \refeq{comp} to \refeq{sB12} determines $ C_{1} $ and leads to the bispectrum 
\be\label{B12sym}
\langle  \varphi(\textbf{k}_1,\tau)\varphi(\textbf{k}_2,\tau)\varphi(\textbf{k}_3,\tau)\rangle'=\frac{\pi H^2V'''(\bar{\phi})}{8}\frac{\tau^3}{k_1 k_2 k_3}\,.
\ee
In the next section, we will confirm both \refeq{B32sym} and \refeq{B12sym}, obtained from symmetries, with an explicit calculation using the in-in formalism. 

Eventually, we are interested in the correlators of curvature perturbations $ \zeta $, which, unlike those of $ \varphi $, are conserved on superHubble scales in single-field inflation. As discussed in \refss{ss:ps}, we have at least two distinct options to convert to $ \zeta $ fluctuations. We can convert at some $ k $-dependent time so that the tilt hides in the time-dependent factors in \refeq{B32sym}. It is evident that the expression simplifies considerably if we convert at ``\textit{perimeter crossing}'' time (p.c.), defined by the solution of
\be
-1+\gamma_E+ \log{(-K \tau_{\text{p.c.}})}=0\,.
\ee
Using the triangle inequality, one finds $ K\geq 2k_{i} $ for every $ i $. In particular $ K\geq 2k_{\text{max}} $ for $ k_{\text{max}} $ the largest of the three momenta. Using this we can check that 
\be
-k_{\text{max}}\tau_{\text{p.c.}}\geq \frac{e^{1-\gamma_{E}}}{2}\simeq 0.8\,,
\ee
which ensures that the conversion is performed after all the three modes have crossed the Hubble radius. For the bispectrum, we need the gauge transformation between $ \varphi $ and $ \zeta $ at second order. This can be computed directly from the gauge transformation, which we review in \refapp{a:secord}, or using the $ \delta N $ formalism, which we review in \refapp{a:last}. Neglecting all spatial derivatives, which are subleading on superHubble scales, the second order gauge transformation is
\beq\label{closer}
\zeta = -\frac{1}{\sqrt{2\e}\Mpl}\varphi + \frac{\eta}{8\e\Mpl^2}\varphi^2+\O(\varphi^{3})\,.
\eeq
Then, the $\zeta$-bispectrum is related to $\varphi$ correlators via 
\beq
\begin{split}
\langle \zeta_{\k_{1}} \zeta_{\k_{2}} \zeta_{\k_{3}}\rangle'&=\frac{-1}{(2\e)^{\frac{3}{2}}\Mpl^3}\Bigg|_{\text{p.c.}}\langle \varphi(\textbf{k}_1,\tau_{\text{p.c.}})\varphi(\textbf{k}_2,\tau_{\text{p.c.}})\varphi(\textbf{k}_3,\tau_{\text{p.c.}})\rangle'+\\
&~~~+\frac{1}{2\e^2\Mpl^4} \frac{\eta}{8}\Bigg|_{\text{p.c.}}\left(\langle \varphi(\textbf{k}_1,\tau_{\text{p.c.}})\varphi(-\textbf{k}_1,\tau_{\text{p.c.}})\rangle'\langle \varphi(\textbf{k}_2,\tau_{\text{p.c.}}) \varphi(-\textbf{k}_2,\tau_{\text{p.c.}})\rangle'+\text{2 perm.}\right)\,,
\label{varphiresult2}
\end{split}
\eeq
The first term in \refeq{varphiresult2} is the connected part of the $\varphi$ correlator and the second term is the contribution from the non-linear gauge transformation between comoving and flat gauge. 

In comoving gauge, interactions arise from different couplings. It is therefore convenient to express all background parameters, for example the potential and its $\phi$-derivatives, in terms of Hubble slow-roll parameters. Starting from \refeq{beom} and taking three $\phi$-derivatives, in the limit \refeq{declimit}, we obtain (see also \refapp{PotinSRpar})
\beq
V'''(\bar{\phi})= \frac{H^2}{\sqrt{2\e}\Mpl}\left[ - \frac{3}{2}\frac{\dot{\eta}}{H}-\frac{\ddot{\eta}}{2H^2}-\frac{\eta\dot{\eta}}{2H}\right] \,.
\label{Vslowroll1}
\eeq
The bispectrum then takes the form
\be\label{Bzeta1}
\langle \zeta_{\k_{1}} \zeta_{\k_{2}} \zeta_{\k_{3}}\rangle &=& \frac{H^4}{16 \e^2\Mpl^4}\Bigg|_{\text{p.c.}} \left[  \frac{\eta_{\text{p.c.}}}{2} \frac{k_1^3 + k_2^3+k_3^3}{k_1^3 k_2^3k_3^3}+ \frac{\dot{\eta}}{2H}\frac{k_1 k_2 k_3-\sum_{i\neq j}k_i^2 k_j }{k_1^3k_2^3k_3^3} \right] \,,
\ee
where we omitted the label p.c. in $ \dot \eta $ since there it leads to a correction that is further suppressed in the slow-roll expansion. It is also convenient to proceed differently and make the momentum dependence explicit. As we did in \refss{ss:ps}, we can convert all modes at some late time $ \tau_{\ast} $ that is the same for every mode. Using the relation \eqref{closer}, we find
\be\label{finalsym}
\langle \zeta_{\k_{1}} \zeta_{\k_{2}} \zeta_{\k_{3}}\rangle &=& \frac{H^4}{16 \e^2\Mpl^4}\Bigg|_{\ast} \Bigg\{\frac{\eta_{\ast}}{2} \frac{k_1^3 + k_2^3+k_3^3}{k_1^3 k_2^3k_3^3}+ \frac{\dot{\eta}}{2H}\frac{1}{k_1^3k_2^3k_3^3}\times \\
&&   \Big[(-1+\gamma_E+ \log{(-K \tau_*)})\sum_{i=1}^3 k_i^3-\sum_{i\neq j}k_i^2 k_j +  k_1 k_2 k_3\Big]\Bigg\}\,, \nonumber
\ee
where now all factors are $ k $-independent. 

As expected, our final expression for the bispectrum obeys the squeezed limit consistency condition of Maldacena \cite{Maldacena:2002vr}. To see this, consider the limit $ k_{1}\equiv k_{L}\ll k_{2,3}\equiv k_{S} $ of \refeq{finalsym}, 
\begin{align}
\label{bispsque}
\langle \zeta_{\k_{L}} \zeta_{\k_{S}} \zeta_{\k_{S}}\rangle & = \left( \frac{H^2}{4 \e\Mpl^2} \right)^{2}  \frac{1}{k_{L}^{3}k_{S}^{3}} \left[ \eta_* + \frac{\dot{\eta}}{H} (-2+\gamma_E+ \log{(- 2k_S \tau_*)})\right] +\O\left(  \frac{k_{L}^{2}}{k_{S}^{2}}\right) \nonumber \\
& =P(k_L)P(k_S) \left[ \eta_* + \frac{\dot{\eta}}{H}\log{(- k_S \tau_*)} + \frac{\dot{\eta}}{H} (-2+\gamma_E+ \log{2})\right] +\O\left(  \frac{k_{L}^{2}}{k_{S}^{2}}\right) \nonumber \\
&= -(n_s-1) P(k_L)P(k_S)\;,
\end{align}
where $P(k)$ is the $\zeta$ power spectrum. Indeed, the spectral tilt to second order in slow-roll is given by \cite{Lyth:1998xn}
\be
n_s-1 = -2\varepsilon -\eta -2\varepsilon^2 - \varepsilon \eta - \frac 1 H (2\dot\varepsilon + \dot \eta)(-2+\gamma_E+ \log{2}) \;,
\ee
which in the limit $ \e\ll \eta $ becomes
\be
n_s-1 = -\eta - \frac{\dot \eta}{H}(-2+\gamma_E+ \log{2}) \;. 
\ee

The extra term in the second line of Eq.~\eqref{bispsque} is due to the time dependence of $\eta$. Its role is to ensure that the scale dependence of $P(k_S)$ is evaluated at the horizon crossing for the short modes. It is easy to see that 
\be
\eta_* + \frac{\dot{\eta}}{H} \log{(- k_S \tau_*)} = \eta(\tau_{\rm H.c.})\;,
\ee
where $\tau_{H.c.}=-1/k_S$. If we set $\tau_* = \tau_{\rm H.c.}$ form the beginning, we immediately get the correct formula for the tilt from the squeezed limit of the three-point function. The explicit check can be found in Appendix \ref{ssec:btw} (see Eq.~\eqref{varphiresult3}).
 
\subsection{Conformal non-Gaussianity}
\begin{figure}
\centering
\begin{tabular}{cc}
\includegraphics[width=0.5\textwidth]{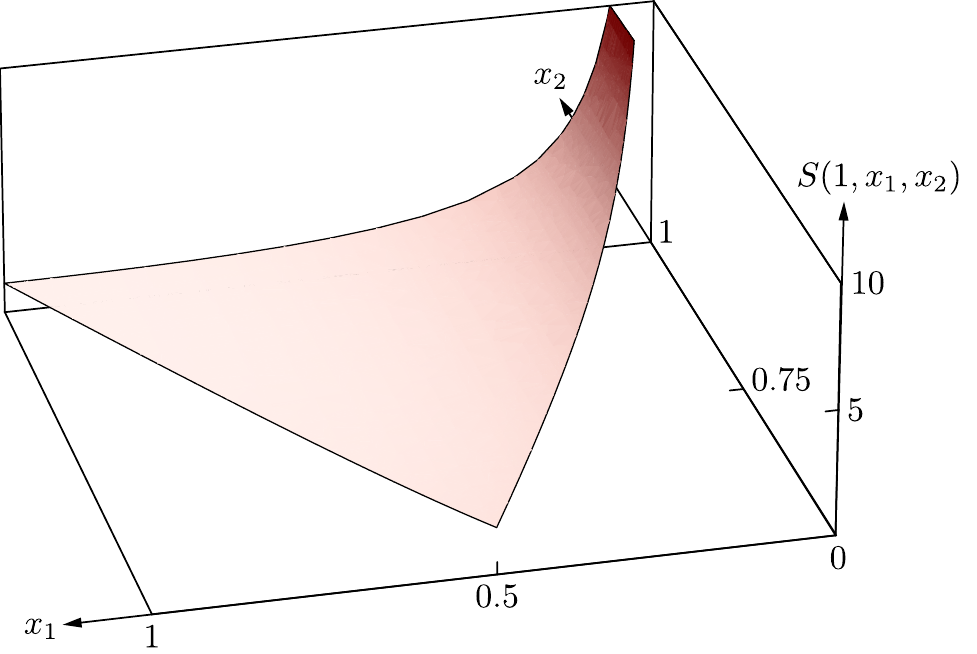} & \includegraphics[width=0.5\textwidth]{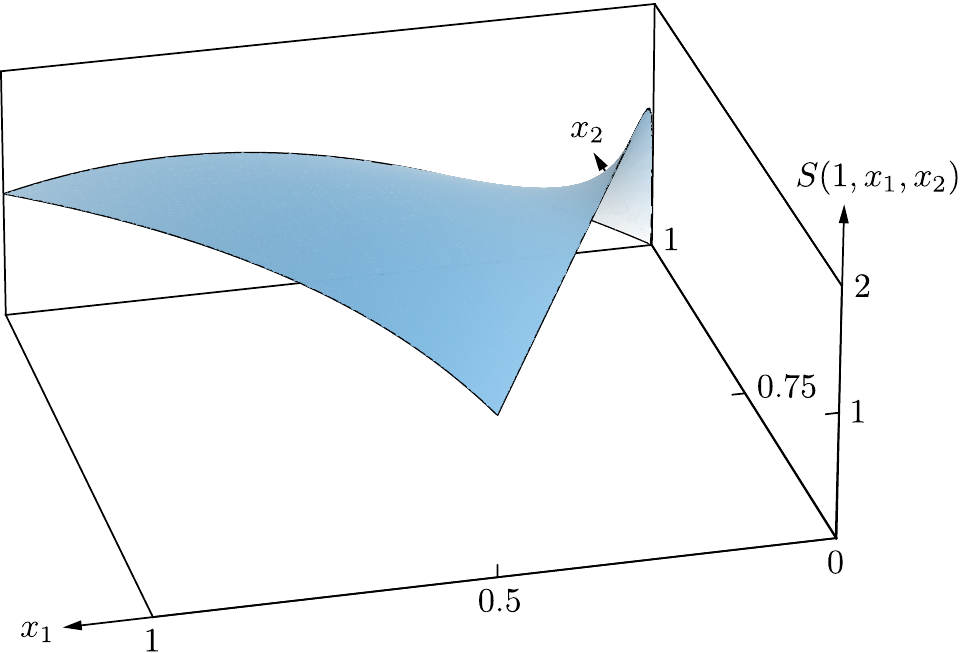} 
\end{tabular}
\caption{\label{shapes} We plot of the shape function $S(k_{1}, x_2, x_3)\equiv (x_2 x_3)^2 \langle \zeta(|\k_1|=1)\zeta(|\k_2|=x_2)\zeta(|\k_3|=x_3)\rangle'$ of the conformal shape for the fundamental domain $1-x_1 \leq x_2 \leq x_1 \leq 1$, with $0.5\le x_2\le 1$. We have chosen $ k_{1} $ (and $ \tau_{\ast}) $ in such a way that the squeezed limit is positive (left hand plot) or vanishes (right hand plot). The figures are normalized to have the value 1 for the equilateral configuration $x_2 =x_3=1$.}
\label{shapeplot}
\end{figure}

Let us finally discuss some phenomenological features of the bispectrum we calculated. Following \cite{Babich:2004gb}, in \reffig{shapes} we plot the fundamental domain of the conformal shape \eqref{equilog} for two different choices of the overall scale (recall that the shape is not scale invariant). For generic value of the overall scale $ k_{1} $, the squeezed limit is positive (left hand plot). The shape is hardly distinguishable from the local shape, \refeq{local}. Both shapes peak in the squeezed limit, and are smooth elsewhere. However, notice that the squeezed limit of the conformal shape switches sign depending on the $ k_{1} $ slice chosen for the plot. In the sense of the cosine defined in \cite{Babich:2004gb}, the conformal shape is therefore orthogonal\footnote{The cosine can be made to vanish by an appropriate choice of $ \tau_{\ast} $, depending on the experiment under consideration.} to the local one and therefore should be searched for separately. Notice that the $\log \times \text{local}$ part becomes very large in the squeezed limit and dominates the conformal shape. Therefore, we also show the plot for the $ k_{1} $ which makes the squeezed limit vanish (right hand plot). This highlights the details of the equilateral configurations.

The conformal shape of the bispectrum in \refeq{equilog} is scale dependent. Therefore, to quantify the size of its subleading contribution to the bispectrum \refeq{finalsym} in terms of an overall size $ f_{NL}^{\rm conf.} $, we need to specify not only the relative size of the wavenumbers but also their specific value. We choose $ k_{1}=k_{2}=k_{3}=k_{p}$ with $ k_{p}=e^{1-\gamma_{e}}/(-3\tau_{\ast}) $, so that the term multiplying  $\sum_i k_i^3$ vanishes. We find
\beq\label{fnldef}
f^{\rm conf.}_{NL}\equiv \frac{5}{18}\frac{\ex{ \zeta\zeta\zeta }_{\rm conf.}}{\langle \zeta \zeta\rangle^2}\Bigg|_{k_{p}}=-\frac{25}{36}\alpha_s \, ,
\eeq
where we used
\be
\alpha_s= \frac{d}{d\log k} n_s \simeq -\frac{\dot\eta}{H}.
\ee
Needless to say, it will be a tall order to measure such a small effect. Nonetheless, the pursuit of non-Gaussianity and the goal to establish single-field inflation might lead us, one distant day, to eventually test \eqref{fnldef}.


\section{The Bispectrum from the In-In Formalism}\label{s:bisp}

In this section, we compute the bispectrum in the conformal limit using the in-in formalism. We show that, in the decoupling limit, \refeq{declimit}, the cubic terms of the $\varphi$ action simplify dramatically. We state the expression for the bispectrum in the spatially flat gauge and work out the transformation of the bispectrum to the comoving gauge. We show that this transformation can be related to the $\delta N$-formalism and relate the running of the scalar tilt, $\alpha_s$, to the $\zeta$-bispectrum. Last, as a consistency check, we obtain the same result directly in the comoving gauge. In passing, we clarify some subtleties of working directly with the $\zeta$ fluctuation, which elucidate the role of boundary terms in the action. 

 
\subsection{Flat Gauge}

To calculate the three-point function for $\varphi$, we use the in-in formalism (see e.g. \cite{Weinberg:2005vy}), with
\beq
\left\langle  {\cal O}(\tau)\right\rangle = \left\langle \left[\bar{T}\exp\left(i \int_{-\infty^{-}}^{\tau}dt~H_{int}(t') \right) \right]{\cal O}^I(\tau)\left[T\exp \left(-i \int_{-\infty^+}^{\tau}dt~H_{int}(t')\right) \right] \right\rangle,
\eeq
where the superscript $I$ indicates that the fields on the right-hand side are in the interaction picture, 
$T$ and $\bar{T}$ refer to time-ordering and anti-time-ordering, and we must perform a suitable $i0$ contour rotation to project onto the interacting vacuum. We will consider ${\cal O}(\tau) = \varphi(\k_1,\tau)\varphi(\k_2,\tau)\varphi(\k_3,\tau)$. Also, as noticed in \cite{Maldacena:2002vr}, at cubic order in perturbations we have $H_{int}=-\mathcal{L}_{int}$. 

To calculate the bispectrum, the action has to be expanded to cubic order in perturbations. For the cubic action, this means that we only need to solve for $N$ and $N_i$ to first order in perturbations\footnote{As shown in \refapp{constrtoorder}, with the solutions of the constraints to the $n^{\rm th}$ order, we can determine the action up to the $(2n+1)^{\rm th}$ order in perturbations. This is a stronger result than the one previously proven in \cite{Chen:2006nt}.}. To emphasize the Planck suppression, we write schematically (see \refapp{a:A} for explicit formulas)
\beq
N=1+\Mpl^{-2} N^{(1)}+\cdots,~~N_i=\Mpl^{-2}N^{(1)}_i +\cdots\, .
\eeq 
Plugging these in the action \eqref{equ:infac} and expanding to cubic order, we obtain 
\be
S_3 &=&\int d^4x~a^3\Bigg\{ -\frac{1}{6}V'''(\bar{\phi})\varphi^3-  \nonumber\\ && -\Mpl^{-2}\Big[\frac{1}{2} V''(\bar{\phi})\,\varphi^2N^{(1)}+\frac{1}{2}  \dot{\varphi}^2 N^{(1)} + \frac{1}{2 a^2} (\partial\varphi)^2 N^{(1)}+ \frac{1}{a^2} \dot{\varphi}\, \partial_i \varphi \, N_i^{(1)}\Big]+  \nonumber\\
&&+\Mpl^{-4} \Big[-\frac{1}{2 a^4}\left( (\partial_i N^{(1)}_j)^2 - (\partial_i N_i^{(1)})^2\right)N^{(1)}-\frac{2H}{a^2}\partial_i N_i^{(1)} (N^{(1)})^2+\frac{1}{a^2} \dot{\bar{\phi}}\, \partial_i\varphi\,N_i^{(1)} N^{(1)}  \nonumber
\\
&&+3H^2\, (N^{(1)})^3 +\dot{\bar{\phi}}\, \dot{\varphi}\,(N^{(1)})^2\Big]-\Mpl^{-6}\, \frac{\dot{\bar{\phi}}^2}{2}(N^{(1)})^3\Bigg\}\,. \label{S3sffull}
\ee
To leading order in the conformal limit, \refeq{declimit}, both $N^{(1)}$ and $N^{i(1)}$ vanish (see \refapp{a:A}), since they are $O(\sqrt{\e}/\Mpl)$; moreover, all the terms in the action that contain $(N^{(1)}, N^{i(1)})$ are $\Mpl$ suppressed. The only nonvanishing term in $S_3$ comes from expanding the potential around the background value of $\phi$,
\beq
S_3 = \int d^4x~a^3\left[-\frac{1}{6}V'''(\bar{\phi})\varphi^3\right].
\label{S3}
\eeq
This is expected since the conformal limit implies the decoupling limit. The Bunch-Davies mode functions for $ \varphi $ are given by (see \refapp{a:ff} for a review of canonical quantization of inflationary fluctuations)
\beq
u_k(\tau) = - e^{-i\frac{\pi}{2}(\nu+ \frac{1}{2})}\frac{\sqrt{\pi}}{2}H (-\tau)^{\frac{3}{2}}\text{H}_{\nu}^{(1)}(-k \tau) \, ,
\label{modefuncsol}
\eeq
where 
\be\label{nu}
\nu \equiv \sqrt{9/4 - V''/H^2}\,,
\ee
and $\text{H}_{\nu}^{(1)}$ is the Hankel function of the first kind.

The three-point function of a scalar field $\varphi$ can be calculated exactly when $\nu$ is half-integer, since then the mode functions \refeq{modefuncsol} are expressible in terms of elementary functions. There are two cases that lie within the range $0<\nu\leq 3/2$, namely $\nu= 3/2$ and $\nu = 1/2$, corresponding to a mass $V''=0$ and $V''= \sqrt{2}H$, respectively. The equal-time, three-point function of a massless scalar field in de Sitter is \cite{Zaldarriaga:2003my, Seery:2008qj} 
\beq
\begin{split}
\langle  \varphi(\textbf{k}_1,\tau)\varphi(\textbf{k}_2,\tau)&\varphi(\textbf{k}_3,\tau)\rangle' =\\
&= \frac{H^2}{k_1^3k_2^3k_3^3}\frac{ V'''(\bar{\phi})}{12}\left[(-1+\gamma_E+ \log{(-K \tau)})\sum_{i=1}^3 k_i^3-\sum_{i\neq j}k_i^2 k_j +  k_1 k_2 k_3\right],
\end{split}
\label{B32}
\eeq
where $\tau$ is an arbitrary late time, at which all the physical wavelengths of the particles are bigger than the Hubble radius. This expression and its conformal invariance have been nicely discussed in \cite{Creminelli:2011mw}, in the context of inflation with a spectator field. Here we notice that this result agrees with our derivation from symmetry only, \refeq{B32sym}. In other words, inflaton perturbations in the conformal limit, and therefore in our universe, behave as perturbations of a spectator field in de Sitter, to leading order in slow-roll, and $ \e\ll \eta $.
To provide another check of our derivation based on symmetries, we quote also the three-point function of a scalar field with $V''=2 H^{2}$ \cite{Creminelli:2011mw}
\beq
\langle  \varphi(\textbf{k}_1,\tau)\varphi(\textbf{k}_2,\tau)\varphi(\textbf{k}_3,\tau)\rangle'=\frac{\pi H^2V'''(\bar{\phi})}{8}\frac{\tau^3}{k_1 k_2 k_3}\,.
\label{B12}
\eeq
Note that such a large mass is not relevant to the inflationary computation - the slow-roll conditions imply that our field $\varphi$ is almost massless, with $V'' \ll H^{2}$. However, the fact that \eqref{B12} agrees with \refeq{B12sym}, provides an additional check of our symmetry arguments in \refsec{s:softth}.

Since \refeq{B32} agrees with what we found in \refss{ssec:together} from symmetries, the derivation of the $ \zeta $ bispectrum proceeds in the same way as in that subsection and gives again \refeq{Bzeta1} (or \refeq{finalsym}).

\subsection{Comoving Gauge}

In this subsection, we calculate the bispectrum directly in comoving gauge. In order to highlight the correct slow-roll suppression of the terms in the cubic $ \zeta $ action, one must integrate by parts \cite{Maldacena:2002vr}. As a consequence, some of the terms in the cubic action have less than two derivatives, naively violating the theorem that guarantees the freeze-out of superHubble $\zeta$ fluctuations\footnote{The violation is most clearly seen from the proof of the tree-level conservation of $ \zeta $ by Maldacena \cite{Maldacena:2002vr}.}. This spurious time dependence is canceled by a suitable boundary term in the action. Moreover, there is a spurious time dependence if one computes the bispectrum using de Sitter mode functions. We clarify how to solve this small puzzle with a toy model, and proceed to do the computation of the bispectrum, finding agreement with the flat gauge calculation.

If we naively calculate the cubic action for $\zeta$, we find it to be $O(1)$ and not slow-roll suppressed, contrary to the quadratic action for $\varphi$, which is $O(\varepsilon)$. In \cite{Maldacena:2002vr}, Maldacena emphasized that the cubic action for $\zeta$ is actually $O(\varepsilon^2)$, in the standard slow-roll expansion. To do that, he integrated by parts the cubic action, to bring it to the form
\begin{equation}
\begin{split}
S_3 =&\Mpl^2 \int \text{dt}~\text{d}^3\text{x} \left[a^3 \e^2 \zeta \dot{\zeta}^2 - 2 a \e^2 \dot{\zeta}(\partial\zeta)(\partial\chi) + a \e^2 \zeta (\partial \zeta)^2 +\right.\\
&\left. +\frac{a^3}{2} \varepsilon \dot{\eta} \zeta^2 \dot{\zeta}  + \frac{1}{2} \frac{\varepsilon}{a} \partial \zeta \partial \psi^{(1)} \partial^2  \psi^{(1)}+ \frac{\varepsilon}{4}\partial^2 \zeta ( \partial \psi^{(1)})^2 + f(\zeta) \frac{\delta \mathcal{L}_2}{\delta \zeta}\right]\,,
\end{split}
\label{Scm3}
\end{equation}
where $\delta \mathcal{L}_2/\delta \zeta$ are the second order equations of motion for $\zeta$, and $f(\zeta)$ is
\beq
f(\zeta)\approx \frac{\eta}{4}\zeta^2 + \frac{1}{H}\dot{\zeta}\zeta + \mathcal{O}(\partial \zeta)\,.
\eeq
To obtain \refeq{Scm3}, multiple integrations by parts were performed and all boundary terms were dropped. The bispectrum was then calculated using the in-in formalism and the mode functions of $\zeta$ were approximated by de Sitter mode functions. The action was written into a more convenient form by performing the field redefinition
\bea
&&\zeta \rightarrow \zeta_n + f(\zeta_n)\,,\label{fieldshift}\quad \text{with}\quad f(\zeta_n)\approx \frac{\eta}{4}\zeta_n^2 + \frac{1}{H}\dot{\zeta}_n\zeta_n + \mathcal{O}(\partial \zeta_n)\,.
\eea
The resulting action is equivalent to \refeq{S3sffull} up to a factor of $\varepsilon$, but the newly defined $\zeta_n$ is not conserved on super Hubble scales. To obtain the $\zeta$-bispectrum, the $\zeta_n$-bispectrum needs to be converted to $\zeta$ in a similar fashion to the $\varphi$ to $\zeta$ gauge transformation. 

In \cite{Arroja:2011yj}, the importance of the boundary terms was noticed, since the results of the calculation in \cite{Maldacena:2002vr} could not be derived from \refeq{Scm3} without the field redefinition \eqref{fieldshift}. The boundary term needed to obtain a consistent result is
\beq
S_{\mathcal{B}}=-\Mpl^2\int \text{dt} \frac{d}{dt}\left(\frac{\varepsilon\eta a^3}{2}\zeta^2 \dot{\zeta}\right)=-\Mpl^2\int d\tau~ \frac{d}{d\tau}\left(\frac{\varepsilon\eta a^2}{2}\zeta^2 \zeta'\right)\,.
\label{etaboundary}
\eeq
In \citep{Burrage:2011hd} higher-order contributions to the bispectrum were calculated without the field redefinition, and it was noticed that unphysical time dependences (divergences) would appear in the $\zeta$-bispectrum. The authors of \citep{Burrage:2011hd} showed that these time dependences were due to relying on the de Sitter approximation for the mode functions, and also to the assumption that the cosmological parameters $H$, $\varepsilon$, $\eta$, etc. had no time variation.

When we analyze the terms in \refeq{Scm3} that are relevant to our decoupling limit \eqref{declimit}, similar issues of time dependence in the bispectrum arise. It is useful to consider a simplified toy model to understand these issues. With this understanding, it is easier to derive the correct bispectrum for $\zeta$, which agrees with \refeq{finalsym}, calculated in $\varphi$ gauge. 


\subsubsection{A toy model}

Let us consider the following toy model action
\beq
S=\Mpl^2\int dt~\varepsilon ~a^3 e^{\zeta} \dot{\zeta}^2+\O(\partial_i \zeta)=\Mpl^2\int dt~\left[\varepsilon a^3  \dot{\zeta}^2+\varepsilon a^3 \zeta \dot{\zeta}^2 + \O(\zeta^4,\partial_i\zeta) \right]\,.
\label{TMA}
\eeq
We omit the $\partial_i\zeta$ terms for simplicity, but they must be present in the quadratic action, as we use de Sitter mode functions to calculate the bispectrum. They must also be present in the cubic action, as the $\zeta$ consistency condition can be phrased as a non-linearly realized residual symmetry of the action. Nonetheless, the issue of spurious time dependence in the bispectrum can be understood just by looking at the cubic term presented explicitly in \refeq{TMA}.

When calculating the bispectrum using the cubic action, one makes a couple of approximations.  First of all, the exact mode functions are proportional to a Hankel function. Integrals over these mode functions simplify considerably when the index of the Hankel functions is half integer. Therefore, the mode functions are often approximated by the de Sitter mode functions, namely 
\be
u_k(\tau) &=& \frac{H}{\sqrt{2k^{3}}} \left(  1+ik\tau\right)e^{-ik\tau}\,, \quad (\nu=3/2)\,.
\ee
The second approximation involves the exact time dependence in $\varepsilon$, $\eta$ and all higher order slow-roll parameters. One assumes that these parameters are approximately constant around Hubble crossing. 

To bring the cubic action to the form \eqref{Scm3}, one needs to integrate by parts. Let us mimic this by rewriting the interaction term in \refeq{TMA} as
\beq
\int dt~\varepsilon a^3 \zeta (\zeta')^2 = \int dt~\left[\partial_{\tau}\left(a^2\varepsilon \zeta^2 \zeta'\right)- a^3 \varepsilon\left(3H \zeta^2 \zeta'+ \frac{1}{a}\zeta (\zeta')^2 + \zeta^2 \partial_{\tau}\left(\frac{1}{a}\partial_{\tau}\zeta\right)+\eta H \zeta^2\zeta'\right)\right]\,.
\label{TMA2}
\eeq
In \cite{Maldacena:2002vr} it was proven that all the terms in the action must have at least two derivatives, which ensures that the $\zeta$ bispectrum become constant at late times. However, \refeq{TMA2} contains multiple terms containing only a single time derivative acting on $\zeta$. The appearance of these terms is, of course, due to the integration by parts. Nonetheless, these integrations by parts were important to show explicitly the slow-roll suppression of the cubic action. If we calculate the bispectrum directly from the cubic term in \refeq{TMA}, the result is time independent, 
\beq
\begin{split}
\langle\zeta^3\rangle' = -\frac{H^4}{32 \Mpl^4\varepsilon^2 k_1^3k_2^3k_3^3} \left[\frac{1}{K}\sum_{i < j}k_i^2 k_j^2 + \frac{k_1 k_2 k_3}{K^2}\sum_{i < j}k_i k_j \right]\,.
\end{split}
\label{LHS}
\eeq
However, if the bispectrum is computed using the term multiplying $\eta$ in \refeq{TMA2}, which contains a single time derivative in $\zeta$, we obtain a spurious time dependence. 
This $\O(\eta)$ term is usually absorbed into $f(\zeta) \frac{\delta \mathcal{L}_2}{\delta \zeta}$ (with the addition of suitable cubic terms with spatial derivatives), since it multiplies the $\zeta$ equations of motion. The contribution to the bispectrum coming from this term is formally zero when evaluated with the quasi-de Sitter mode functions. Due to our approximation of using de Sitter mode functions this is no longer true. More importantly, terms where a time derivative has been partially integrated on to a slowly-varying parameter produce a time dependent contribution to the bispectrum. In other words, even though we count $\dot\zeta \zeta^2$ as containing a single time derivative, in some sense we have time derivatives hidden in the slow-roll parameters. To deal with these time dependences, we can Taylor expand the slow-roll parameters around the time $t_*$ when the perturbations become superHubble,
\beq
\begin{split}
\varepsilon(t) \approx\varepsilon(t_*) + (t-t_*) \dot{\varepsilon}(t_*)+~...= \varepsilon_* - \log\left(\frac{\tau}{\tau_*} \right) \varepsilon_* \eta_*  + ~...\,.
\label{SVP}
\end{split}
\eeq
The contribution to the bispectrum coming from the term in \refeq{TMA2} multiplying $\eta$ is 
\beq
\langle \zeta^3\rangle \supseteq \frac{H^4 \eta_* }{32 \Mpl^4 \varepsilon^2 k_1^3 k_2^3 k_3^3}\left( (-1+ \gamma_E + \log(K \tau))\sum_{i=1}^3 k_i^3 - \sum_{i\neq j} k_i k_j^2 + k_1 k_2 k_3 \right)\,.
\label{TMResulta}
\eeq
The additional contribution to the bispectrum that is produced by substituting \refeq{SVP} into the total derivative term of \refeq{TMA2} is
\beq
\begin{split}
\langle \zeta^3\rangle \supseteq -\frac{H^4 \eta_* }{32 \Mpl^4\varepsilon^2 k_1^3k_2^3k_3^3}\log\left(\frac{\tau}{\tau_*}\right)\sum_{i=1}^3 k_i^3\,.
\label{TMResult}
\end{split}
\eeq
After adding \eqref{TMResulta} and \eqref{TMResult}, one finds that the time dependences cancel. Notice that this result is $\O(\eta)$, which is a slow-roll correction to the bispectrum \eqref{LHS}. Nonetheless, by treating properly the boundary term and the time dependence of $\varepsilon$ and $\eta$, this correction to the bispectrum is time independent.


\subsubsection{Bispectrum from $\dot\eta$}
In the conformal limit, \refeq{declimit}, the only non-zero contributions to the bispectrum are those from the term proportional to $\dot{\eta}$ in \refeq{Scm3} and the boundary term \refeq{etaboundary}. Following the logic of the toy model, the leading contributions to the bispectrum, together with the boundary term contribution that exactly cancels the spurious time dependence in the bispectrum, are 
\beq
\begin{split}
&  \langle \zeta(\textbf{k}_1,0) \zeta(\textbf{k}_2,0) \zeta(\textbf{k}_3,0)\rangle'\supseteq\\
&\supseteq\frac{H^4}{16 \Mpl^4\varepsilon^2 k_1^3k_2^3k_3^3}\frac{\dot{\eta}_*}{2 H}\left[ (-1+ \gamma_E + \log(-K \tau))\sum_{i=1}^3 k_i^3 - \sum_{i\neq j} k_i k_j^2 + k_1 k_2 k_3 \right],
\end{split}
\label{etadotresult2a}
\eeq
and
\beq
\begin{split}
& \langle \zeta(\textbf{k}_1,0) \zeta(\textbf{k}_2,0) \zeta(\textbf{k}_3,0)\rangle'\supseteq \frac{H^4  }{16 \Mpl^4\varepsilon^2}\left[ \eta_* - \frac{\dot{\eta}_*}{2H_*}\log\left(\frac{\tau}{\tau_*} \right)\right]\frac{k_1^3+k_2^3+k_3^3}{k_1^3k_2^3k_3^3}\,.
\end{split}
\label{etadotresult2b}
\eeq
Adding \eqref{etadotresult2a} to \eqref{etadotresult2b}, we reproduce \refeq{finalsym}. Note that we can get corrections to the bispectrum that are higher order in slow-roll by integrating by parts the $\dot{\eta}$ term in \refeq{Scm3} once more. 


\section{The Bispectrum from the Wave Function of the Universe}\label{sec:MMM}

We derive the bispectrum using the Wheeler-de Witt formalism and the wave function of the universe. The results of this section were derived in collaboration with Mehrdad Mirbabayi and Marko Simonovi\'c. 

 
\subsection{The Wave Function of the Universe}

The initial condition of our universe can be described by a wave function \cite{Hawking}. The norm-squared of the wave function $|\Psi|^2$ gives the probability distribution for various field configurations at a given time-slice, and equal-time correlation functions are evaluated via a functional integral of schematic form
\be\label{pathint}
\expect{O_1(t,\x_1)\cdots O_N(t,\x_N)}= \int D\sigma |\Psi[t,\sigma]|^2 O_1(\x_1)\cdots O_N(\x_N),
\ee
where $O$'s are local observables, and $\sigma$ stands for all degrees of freedom. Even though in cosmology we measure classical correlations, it is often useful, both conceptually and technically, to work with the wave function. Most relevant to our discussion is the fact that symmetries are realized more directly.\footnote{An interesting example in which the gauge fixing procedure, necessary to evaluate the functional integral in \eqref{pathint}, is incompatible with conformal symmetry of the wave function and hence leads to a non-conformal 4-point correlation function has been studied in \cite{Trivedi}.} 

In perturbation theory, the wave function is expanded in powers of fluctuations as
\be
|\Psi|^2 = \exp\left[\sum \frac{1}{N!}\int_{\{\k_i\}}\Gamma_N(\k_1,\cdots,\k_{N-1}) \sigma(\k_1)\cdots \sigma(\k_N)
(2\pi)^3 \delta^3\left(\sum_i \k_i\right)\right].
\ee
To calculate the correlation functions one can use Feynman rules for a 3d euclidean QFT, with the power spectrum 
\be
\expect{\sigma(\k) \sigma(-\k)}' = P(k) = \Gamma_2^{-1}(k),
\ee
playing the role of the propagator. In particular, the 3-point function is given by
\be
\expect{\sigma(\k_1)\sigma(\k_2)\sigma(\k_3)}' = \frac{\Gamma_3(\k_1,\k_2)}{\Gamma_2(k_1)\Gamma_2(k_2)\Gamma_2(k_3)}.
\ee

In gravity, the wave function must be invariant under diffeomorphisms. This is guaranteed by the Wheeler-de Witt equations, consisting of one Hamiltonian constraint and three momentum constraints \cite{DeWitt}. In single-field inflation they read (we restore $\hbar$ and follow the notation and conventions of \cite{Pimentel})
\bea
\label{H}
\left[\frac{\hbar^2}{2 \kappa \sqrt{h}}G_{ij,kl}{\delta\over \delta h_{ij}}{\delta \over \delta h_{kl}}+\kappa\sqrt{h}R +\frac{\hbar^2}{4\kappa\sqrt{h}}{\delta^2\over\delta\phi^2}-\kappa\sqrt{h}\left(h^{ij}\partial_i\phi\partial_j\phi+2V(\phi)\right)\right]\Psi &= &0 \, ,\\[10pt]
\label{P}
-2 i \hbar \nabla_i\left[\frac{1}{\sqrt{h}}\frac{\delta \Psi}{\delta h_{ij}} \right]+i\hbar{1\over\sqrt h}h^{ij}\partial_i\phi\frac{\delta \Psi}{\delta \phi}&=&0 \, ,
\eea
with $\kappa \equiv \Mpl^2/2$, $h_{ij}$ the spatial metric and $h^{ij}$ its inverse. The momentum constraints ensure the invariance under diffeomorphisms within the 3-dimensional time-slice. As argued in \cite{Pimentel}, invariance under asymptotic spatial diffs, supplemented with a few continuity assumptions, yields an infinite number of single-field squeezed limit consistency conditions, in agreement with the original result of \cite{Hinterbichler}. The Hamiltonian constraint ensures invariance under the choice of the time-slice. Therefore, it should be viewed as a time-evolution equation, but the coordinate time $t$ has to be understood as parameterizing a physically measurable (though not necessarily locally measurable) quantity such as the spatial average of a rolling scalar field $\bar\phi$ or global temperature.  

At superHubble scales, the fluctuations become approximately classical and we can apply the WKB approximation to the Hamiltonian constraint to derive a useful equation for $|\Psi|$. First, write $\Psi = \exp(W/\hbar)$ and note that the variation of the phase $W_i = \rm{Im}\,W$ is rapid, therefore in expectation values the momentum operators is dominated by
\be
\pi_\sigma = - i \hbar \frac{\delta}{\delta \sigma}\approx \frac{\delta W_i}{\delta \sigma},
\ee
and this must agree with the classical value of the momenta (see e.g. \cite{Pimentel}):
\be
\frac{\delta W_i}{\delta h_{ij}} = \kappa \sqrt{h}(E^{ij}-h^{ij}E)= -2\kappa H\sqrt{h} h^{ij} (1+ \O(a^{-2}))
\ee
and 
\be
\frac{\delta W_i}{\delta \phi} = \frac{2\kappa \sqrt{h}}{N} (\dot\phi - N^i\der_i\phi)= 2\kappa \sqrt{h} \dot\phi
(1+ \O(a^{-2})),
\ee
where $E_{ij}$ is the extrinsic curvature and in the final expressions we used the fact that $N = 1 +\O(a^{-2})$ and that $N_i$ is not growing with $a$. Plugging these back in the leading real component of \eqref{H} reproduces the Friedmann equation. Note, however, that the $h_{ij}$ and $\phi$ which appear in the above approximate equations are the full fields including the superHubble fluctuations. Therefore, what we really get is a separate universe version of the Friedmann equation which relates the expansion rate $H$ at the time when $\phi=\bar\phi+\varphi$ to the energy density. The imaginary component of \eqref{H} gives
\be\label{Weyl}
2 H h_{ij} \frac{\delta W_r}{\delta h_{ij}} + (\dot{\bar\phi}+\dot\varphi) \frac{\delta W_r}{\delta \varphi} = 0,\quad  \text{with} \quad W_r = \rm{Re}\,W\,.
\ee
In the exact de Sitter limit where we set $\dot{\bar\phi} =0$ and put a hard cosmological constant, on superHubble scales we have $\dot\varphi = - H \Delta_- \varphi$, with $\Delta_- = 3/2 -\sqrt{9/4 -V''/H^2}$. Then the above equation ensures that the wave function has the right behavior under Weyl transformation to be the generating functional of a $3d$ conformal field theory, with $\varphi$ playing the role of the source for an operator of dimension $\Delta_+ = 3-\Delta_-$. In the same limit, the quadratic and cubic terms in $\varphi$, namely $\Gamma_{\varphi\varphi}$ and $\Gamma_{\varphi\varphi\varphi}$, are fully fixed by conformal symmetry up to three numbers: the amplitude of the 2-point function, and the two coefficients of the local and the conformal shapes as discussed in section \ref{s:softth}.

During inflation and more generally when there is appreciable time derivatives of parameters and hence a preferred choice of time-slicing, the above properties will be lost at sufficiently high order in slow-roll parameters. For instance, if $\dot \eta /H  = \O(\eta^2) \neq 0$ there is a nonzero running which means that $\Gamma_{\varphi\varphi}$ no longer has the required power-law behavior of a conformal theory. On the other hand, since the interactions of $\varphi$ start from $V'''\propto \dot\eta$, at this order $\varphi$ can be considered as a massless field and $\Gamma_{\varphi\varphi\varphi}$ is constrained to be a linear combination of the two shapes \eqref{local} and \eqref{equilog}. Fixing the coefficient to \eqref{Cs} requires the knowledge of the initial condition which is assumed to be Bunch-Davies, and the easiest way to do it is to consider the squeezed limit of $\varphi$ bispectrum as we did in section \ref{s:softth}.

 
\subsection{Curvature Correlators and Gauge Transformations}

Let us now discuss the $\zeta$ correlation functions. In section \ref{s:softth}, we used the time-diffeomorphism that takes us from the flat gauge with $\zeta=0$ to the comoving gauge with $\varphi=0$ to relate correlation functions of $\zeta$ to those of $\varphi$. Below, we will show that one could alternatively use the Wheeler-de Witt equation to relate the probability distributions $|\Psi[\zeta,\varphi=0]|^2$ to $|\Psi[\zeta=0,\varphi]|^2$.

We use $2h_{ij}\delta/\delta h_{ij} = \delta/\delta\zeta$ to write \eqref{Weyl} as
\be\label{dN}
\left[\frac{\delta}{\delta\zeta}- \frac{\dot{\bar\phi}+\dot\varphi}{H}\frac{\delta}{\delta\phi}\right] 
|\Psi[\varphi,\zeta]|^2 = 0.
\ee
As we will now explain, this equation expresses the gauge symmetry of $|\Psi[\varphi,\zeta]|^2$ under the choice of time slicing, which has to be fixed in order to evaluate the functional integral in \eqref{pathint}. In section \ref{s:softth}, we discussed correlation functions in two special gauges, the flat gauge with $\zeta=0$, and the comoving gauge with $\varphi =0$, and a time-diffeomorphism that takes us from one to the other. This diffeomorphism gives a relation $\zeta(\varphi)$ which can be efficiently calculated to any desired order in $\varphi$ using the $\delta N$-formalism:
\be\label{zeta}
\left.\zeta = -\sum_n \frac{1}{n!} \frac{\der^n N}{\der \phi^n}\right|_{\bar \phi} \varphi^n.
\ee
However, there is a continuum of time-slices, generically with $ \zeta\neq0 $ and $ \varphi\neq0 $, which are labeled by different values of $\zeta$. One could perform an infinitesimal time-diffeomorphism with an arbitrary infinitesimal function $\delta\varphi(\x)$ and with $\delta \zeta(\x)$ given by
\be\label{infinitesimal}
\left.\delta \zeta = \frac{\der N}{\der \phi}\right|_{\bar\phi +\varphi} \delta\varphi 
\left. = \frac{H}{\dot{\bar\phi}+\dot\varphi}\right|_{\bar\phi +\varphi} \delta \varphi.
\ee
Equation \eqref{zeta} is an integrated version of this (i.e. a finite transformation) from $[\zeta\neq 0,\varphi=0]$ gauge to $[\zeta=0,\varphi \neq 0]$ gauge:
\be
\left.\zeta = -\int_0^\varphi \frac{\der N}{\der \phi}\right|_{\bar\phi+\varphi'} d\varphi',
\ee
while equation \eqref{dN} ensures the symmetry under the infinitesimal transformation \eqref{infinitesimal}.

Therefore, to find the $\zeta$ bispectrum in the limit $\eps\to 0$ and given the knowledge of $\Gamma_{\varphi\varphi}$ and $\Gamma_{\varphi\varphi\varphi}$, we can alternatively use in \eqref{dN} the fact that $H$ is $\varphi$-independent in this limit and
\be
\dot\varphi = \frac{1}{2}H\eta \varphi +\O(\varphi^2)
\ee
to fix the $\zeta$-dependence of $|\Psi[\zeta,\varphi]|^2$ up to $\O(\zeta^3)$ and then set $\varphi=0$. There is hardly any difference compared to changing the variables at the level of correlation functions and one recovers \eqref{finalsym}. 

For higher-order statistics, it is important to note that the factor $(\dot{\bar\phi}+\dot\varphi)/H$ in \eqref{dN} is generically a nonlinear expression in terms of $\varphi$ because both $\dot\varphi$ and $H$ depend nonlinearly on $\varphi$. It can be calculated perturbatively by inverting
\be
\left.\left.\frac{\der N}{\der \phi}\right|_{\bar\phi+\varphi} = \sum_{n=0}^\infty \frac{1}{n!}\frac{\der^{n+1} N}{\der\phi^{n+1}}\right|_{\bar\phi}\varphi^n.
\ee
However in the limit in which all slow-roll parameters except $\eta$ are sent to zero, equation \eqref{dN} greatly simplifies since in this limit $H$ is independent of $\varphi$ and $\dot\varphi = \frac{\eta}{2}H\varphi$ with no higher order corrections. As a consistency check, it can be easily verified that
\be
\left.\Mpl^n\frac{\der^n N}{\der \phi^n}\right|_{\bar\phi} = \frac{1}{\sqrt{2\eps}}(n-1)!\left(\frac{-\eta}{2\sqrt{2\eps}}\right)^{n-1},
\ee
giving
\be
\left.\frac{\der N}{\der \phi}\right|_{\bar\phi+\varphi} = \frac{H}{\dot \phi}= \frac{H}{\dot{\bar\phi} +\frac{\eta}{2}H\varphi},
\ee
as it should.


\section{Discussion}\label{disc}

\begin{figure}
\centering
\includegraphics[width=.5\textwidth]{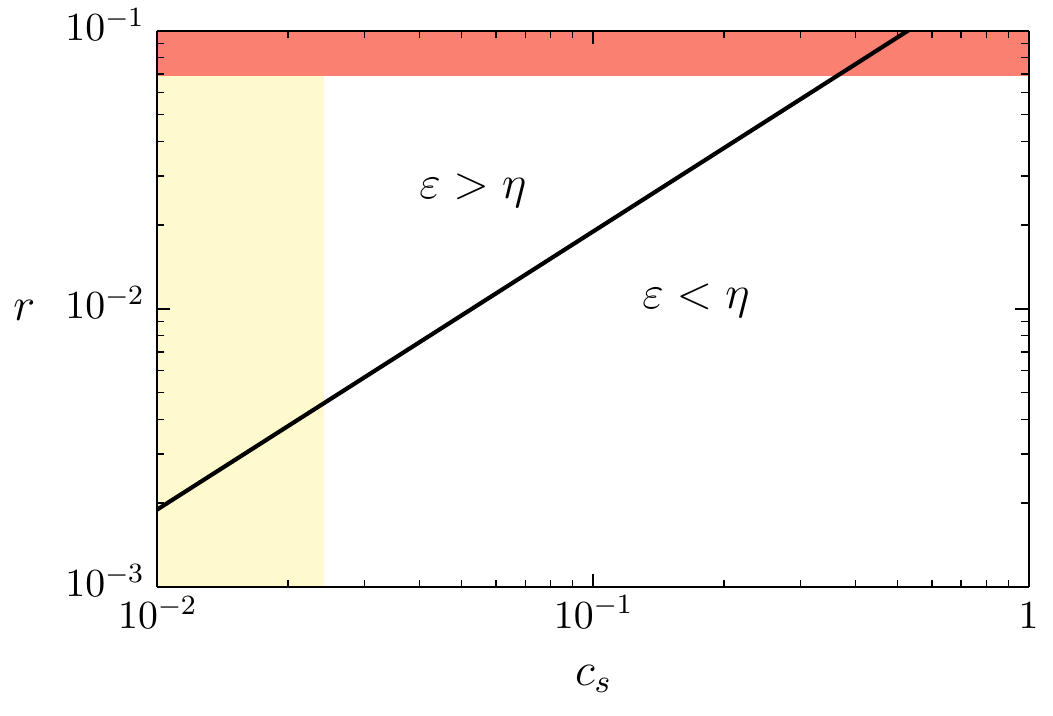}
\caption{Schematic plot of the current experimental constraints on the tensor to scalar ratio $r$ and the scalar speed of sound $c_s$. As the bound on $ r $ and $ c_{s} $ (from $ f_{NL}^{\rm eq.} $) improves, we will exclude the region where $ \varepsilon>\eta $.}\label{figrcs}
\end{figure}

In this paper, we pointed out that, within single-field slow-roll inflation, the non-detection of tensor modes implies $\varepsilon\ll\eta$. We elucidate the consequences of this new hierarchy, which we dub the {\it conformal limit} of inflation: all primordial correlators are constrained by conformal symmetry (a.k.a. de Sitter isometries). Because of the emergent conformal symmetry and with the aid of a new consistency relation, we were able to fully determine the shape and amplitude of the bispectrum as well as the tilt of the power spectrum. Within single-field slow-roll inflation, the predicted size of non-Gaussianity is very small. Our approach relies on the power of symmetries to constrain primordial correlators.

The conformal limit is based on the observation that, given the current CMB bounds, the level of tensor fluctuations is small, and might well turn out to be beyond the reach of any human detection apparatus. In this case, the single-field consistency relation, $r=-8n_t$, might be impossible to test. Nonetheless, another single-field consistency relation arises in the scalar sector as well \cite{Maldacena:2002vr}. Here we have shown that, beyond the well-known local non-Gaussian shape with coefficient $ (n_{s}-1) $, there is a subleading conformal shape, whose amplitude is fixed as well, this time in terms of the running $\alpha_s$. While both these signals are challengingly small, they might well be enormously larger than any tensor observables, for which our theoretical prior extends down to $ r\sim 10^{-55} $.

The slow-roll parameter $\varepsilon$ is fixed by the tensor-to-scalar ratio $r$ only in slow-roll single-field inflation with a canonical kinetic term. In almost any deviation from this minimal model (such as multifield inflation or $P(X)$ theories) it is possible to produce unobservable tensor modes while keeping $\varepsilon \sim \eta$. However, at the same time, this leads to observable non-Gaussianity. Therefore, by producing stronger bounds on $f_{\rm NL}$ and $r$, future experiments will be able to unequivocally establish that the hierarchy $\varepsilon\ll\eta$ is realized in our universe. As an example, let us consider inflationary models with small speed of sound $c_s$. In the plot of \reffig{figrcs}, we show the current bounds on $r$ (from searches for B-mode polarization in the CMB) and $c_s$ (from constraints in non-Gaussianity) in the shaded regions, and divide between the $\varepsilon >\eta$ and $\varepsilon<\eta$ scenarios. To draw \reffig{figrcs}, we use the current measurement of $n_s-1$ as a constraint on $2\varepsilon+\eta$. Future experiments, such as COrE \cite{Bouchet:2011ck}, will be able to rule out entirely the $\varepsilon>\eta$ region, by decreasing the upper bound on $r$. As the experimental bounds become more stringent, the current hierarchy $\varepsilon<\eta/6$ for $c_s=1$ can increase more than ten-fold, confirming our working assumption that $\varepsilon<\eta^2$, such that the bispectrum that we calculate in this paper is the dominant one.

One might be worried that the amplitude of the bispectrum of order $\mathcal O(\eta^2)$ is so small that any correction to the minimal inflationary scenario (such as derivative interactions or heavy fields) would spoil the predictions. While this might be the case in some scenarios, there are natural models in which the bispectrum we calculate in this paper is parametrically larger than all other contributions. As a concrete example, let us consider a Starobinsky-like inflation with the potential \refeq{staro}. Without fine-tuning, derivative operators such as $(\partial \phi)^4/M^4$ must be present and, if $M$ is small, they lead to enhanced equilateral non-Gaussianity. We can estimate this as \cite{Cheung:2007st}
\be
\left. f_{\rm NL}^{\rm eq.} \zeta \sim \frac{\mathcal{L}_3}{\mathcal{L}_2}\right|_{k/a \sim H}.
\ee
Using $\mathcal{L}_3 \sim \frac{\dot{\bar\phi}}{M^4} \dot\varphi^3$, $\zeta \sim H^2/{\dot{\bar\phi}}$ and $\dot{\bar\phi}^2 = 2\e \Mpl^2 H^2$, we get
\be
f_{\rm NL}^{\rm eq.} \sim \frac{\dot{\bar \phi}^2}{M^4} \sim \frac{1}{N^2} \left( \frac H M \right)^2 \sim \eta^2 \left( \frac H M \right)^2 \;,
\ee
where we used 
\be\label{epeta}
\varepsilon \simeq \left( \frac M{M_{\rm Pl}} \right)^2 \frac{2}{N^2} \simeq \left( \frac M{M_{\rm Pl}} \right)^2 \frac{\eta^{2}}{2}  \;.
\ee
Therefore, as long as $H\ll M \ll M_{\rm Pl}$, the bispectrum up to next-to-leading order is the one calculated in this paper. The same argument is expected to apply to all models with small field range $M$, as long as the scale suppressing derivative interactions is of order of or larger than $M$. In this case the first estimate in \eqref{epeta} is a consequence of the Lyth bound \cite{Lyth:1996im}, and $1/N<\eta $ is an empirical constraint.
\\ \\

\noindent{\bf Acknowledgments} We are grateful to Mehrdad Mirbabayi and Marko Simonovi\'{c} for collaboration during various stages of this project. We also thank Daniel Baumann, Giovanni Cabass, Garrett Goon, Hayden Lee and Yvette Welling for discussions and comments on a draft of this paper. A special thanks to Hayden Lee for help with producing the plots in \reffig{shapeplot} and \reffig{figrcs}. E.P. is supported by the Delta-ITP consortium, a program of the Netherlands organization for scientific research (NWO) that is funded by the Dutch Ministry of Education, Culture and Science (OCW). G.P. acknowledges support from a Starting Grant of the European Research Council (ERC STG Grant 279617). 
\appendix

 
\section{Solution of the Constraint Equations}\label{a:A}

In this appendix, we make some comments on the solutions to the constraint equations for the non-dynamical components of the metric. If we are only interested in scalar fluctuations, in the decoupling limit, it turns out to be simpler to work in flat gauge $ \varphi\ne 0 $, $ \zeta=0 $. In fact, we will show that, in our decoupling limit, the constraint equations have trivial solutions, in the sense that
\bea\label{solconst}
N=1+O(\Mpl^{-2})\,  \textrm{and}\,\, \,\,N^i=O(\Mpl^{-2})\, .
\eea

If we insist in working in comoving gauge $ \varphi= 0 $, $ \zeta\ne0 $, the constraint equations simplify very little in the decoupling limit. The best one can do is to discard terms of $O(\varepsilon)$ from the constraint equations. To reconcile the existence of the trivial solution \refeq{solconst} with the complicated expressions in comoving gauge, we work out the coordinate transformations that takes one from flat to comoving gauge. Most of the structure of the solutions of the constraint equations are simply a reflection of a change of coordinates, which is quite remarkable. This approach can be useful in solving the constraint equations, as it is easier to work out the explicit change of coordinates from $\varphi$ to $\zeta$ gauge.


\subsection{Flat Gauge}

Let us recall a few standard facts about the ADM decomposition of the metric. We take the metric on the spatial slice $h_{ij}$ as our data, and solve for the lapse function and shift vector $(N,N^i)$ as a function of $h_{ij}$. The lapse and shift appear with no time derivatives in the action, and thus can be integrated out, as all they do is to impose certain constraints in the spacetime metric $g_{\mu\nu}$, given the metric $h_{ij}$. The metric is
\beq\label{admmet}
\dd s^2=-N^2 \dd t^2+h_{ij}(N^i \dd t+ \dd x^i)(N^j \dd t+\dd x^j)~~.
\eeq

In flat gauge, the scalar fluctuations are contained in the field $\phi$, while the tensor fluctuations remain in the metric $h_{ij}$,
\beq
\phi=\bar\phi(t)+\varphi(\x,t),~~~~h_{ij}=a(t)^2(e^\gamma)_{ij}\equiv a(t)^2\left(\delta_{ij}+\gamma_{ij}+\frac{1}{2}\gamma_{ik}\gamma_{kj}+\cdots\right) ,
\eeq
with $\partial_i\gamma_{ij}=0$ and $\gamma_{ii}=0$. Now write the action (\ref{equ:infac}) in terms of the fluctuations and Lagrange multipliers. We obtain
\bea\label{acconst}
S&=&\int \dd t\dd ^3x\, N a^3 \left( \frac{\Mpl^2}{2} R+\frac{1}{2}\left(N^{-2}(\dot\phi-N^i\partial_i\phi)^2-h^{ij}\partial_i\phi\partial_j\phi\right)-V(\phi)\right)=\\&=&\int \dd t \dd^3 x\, a^3 \left[\frac{\Mpl^2}{2}\left(NR^{(3)}+N^{-1}h^{ij}h^{kl}(E_{ik}E_{jl}-E_{ij}E_{kl})-6NH^2 \right)+\right.\nonumber\\&&+\left.\frac{1}{2}\left(N^{-1}(\dot{\bar\phi}+\dot\varphi-N^i\partial_i\varphi)^2-Nh^{ij}\partial_i\varphi\partial_j\varphi\right)+\right.\nonumber\\&&\left.+N\frac{\dot{\bar\phi}^2}{2}-N\left(V'(\bar\phi)\varphi+\frac{V''(\bar\phi)}{2}\varphi^2+\cdots\right)\right]\nonumber,
\eea
with 
\beq
E_{ij}\equiv N K_{ij} =\frac{1}{2}\left(\dot h_{ij}-\nabla_iN_j-\nabla_jN_i\right),
\eeq
$K_{ij}$ being the extrinsic curvature, and $R^{(3)}$ being the curvature of the spatial slice. 

In the second equality of (\ref{acconst}), we explicitly separated the ``large" piece of the action, with a prefactor $\Mpl^2$, and the ``small" piece of the action, with no $\Mpl^2$ prefactor. When we expand the potential around the background solution, a piece of its background value $V(\bar\phi)$ is of $O(\Mpl^2)$, and sets the cosmological constant, while the subleading piece is proportional to the kinetic term $\sim \dot{\bar\phi}^2$, and thus carries no $\Mpl^2$; we made this separation explicit in the last line of (\ref{acconst}). Finally, the other terms in $V(\phi)$ involve fluctuations and derivatives of the potential. In our decoupling limit, these derivatives are finite, and thus belong to the ``small" piece of the action.

A few comments are in order. Notice that the large piece of the action has no dependence on the scalar sector. This is intuitively clear. In the limit $ \Mpl\rightarrow \infty $ the matter content of the theory, in this case the scalar field, does not gravitate. Two lumps of $ \phi$ do not attract each other to leading order in this limit. Geometry has become a fixed, non-trivial background on which $ \phi $ excitations propagate. The metric is still a dynamical field since gravitons can propagate, but they do not interact with $ \phi $ to leading order (i.e. for $ \Mpl\sim \infty $).

Now suppose that we discard the graviton degrees of freedom, setting $h_{ij}=a^2\delta_{ij}$. The constraint equations read
\bea
&&\Mpl^2\left(R^{(3)}-6H^2-N^{-2}h^{ij}h^{kl}(E_{ik}E_{jl}-E_{ij}E_{kl})\right)-\nonumber\\&&-\Mpl^0\left(N^{-2}(\dot{\bar\phi}+\dot\varphi-N^i\partial_i\varphi)^2+(-\dot{\bar\phi}^2+2V'(\bar\phi)\varphi+\cdots )+h^{ij}\partial_i\varphi\partial_j\varphi \right)=0\,,\\
&&\Mpl^2\left(\nabla_a\left(N^{-1}(h^{ab}E_{bi}-\delta^a_i h^{bc}E_{bc})\right)\right)+\Mpl^0\left( N^{-1}\partial_i\varphi(N^j\partial_j\varphi-\dot{\bar\phi}-\dot\varphi)\right)=0\,.
\eea

As we pointed out before, the leading order terms in the constraint equations have no explicit $\varphi$ dependence, so we expect that they are solved self-consistently at leading order in $\Mpl^2$. It is easy to check that $N=1$ and $N^i=0$ solve the constraint equations, up to terms of $O(\Mpl^{-2})$. We can see that, to self-consistently solve the constraint equations, one must employ the following ansatz:
\bea
N&=&\Mpl^0 (1+O(\gamma^2)) + \Mpl^{-2}O(\varphi, \gamma^2)+O(\Mpl^{-4})\\
N^i&=&\Mpl^0 (O(\gamma^2)) + \Mpl^{-2}O(\varphi, \gamma^2)+O(\Mpl^{-4})
\label{ansatz}
\eea
The graviton appears even in the strict $\Mpl^2 \to \infty$ limit, because we recover the limit of pure gravity in de Sitter space, and the constraint solutions in this case are still non-trivial. There are no linear terms in the graviton fluctuations due to our gauge fixing conditions of transversality and tracelessness. The $O(\Mpl^{-2})$ piece is determined by successive approximations - we evaluate the $\Mpl^2$ term in the constraint equations to $O(\Mpl^{-2})$, and the $\Mpl^0$ term in the constraint equations to $O(1)$, etc. In fact, the explicit calculation to linear order in perturbations shows \cite{Maldacena:2002vr}
\be
N=1+\Mpl^{-2}\frac{\dot{\bar\phi}}{2H}\varphi+O(\varphi^2)\,,\quad N^{i}=-\Mpl^{-2}\frac{\dot{\bar\phi}}{2H}\frac{\partial_{i}}{\partial^{2}}\left[  \dot\varphi+\varphi H \frac{\eta}{2}\right]+O(\varphi^2)\,.
\ee
 
There is an interesting simplification of the constraint equations in the scalar sector, if we use $\Mpl^{-2}$ as a small expansion parameter, rather than $\varphi$. In other words, consider setting $\gamma_{ij}=0$, keeping all powers and derivatives of $\varphi$, but expand the lapse and shift to $O(\Mpl^{-2})$, 
\bea
N(\varphi,\gamma=0)&=&1+\Mpl^{-2}N^{(1)}(\varphi)+O(\Mpl^{-4})\nonumber \\ N_i(\varphi,\gamma=0)&=&\Mpl^{-2} N^{(1)}_i(\varphi)+O(\Mpl^{-4}) \, .
\eea
It will be convenient to use the Helmholtz decomposition of a vector field, $N^{(1)}_i=\partial_i\psi^{(1)}+\widetilde N^{(1)}_i$. Notice that, in the flat gauge, the geometric quantities simplify considerably,
\bea
R^{(3)}=0,\, E_{ij}=Hh_{ij}-\frac{1}{2}(\partial_iN_j+\partial_jN_i)\, \, .
\eea
Now we solve the constraint equations perturbatively in $\Mpl^2$. To the order we are interested in, the constraint equations read
\bea\label{consmpt}
&6HN^{(1)}+a^{-2}\partial_iN_i^{(1)}+L=0 \, \nonumber,\\
&2H\partial_iN^{(1)}+\frac{a^{-2}}{2}\left(\partial_i\partial_j N_j^{(1)}-\partial^2N_i^{(1)}\right)+P_i=0 \, ,
\eea
with
\bea
&L&\equiv\frac{1}{H}\left(\dot{\bar\phi}\dot\varphi+\frac{1}{2}\dot\varphi^2+\frac{1}{2}h^{ij}\partial_i\varphi\partial_j\varphi+(V'(\bar\phi)\varphi+\cdots)\right)\, , \nonumber\\
&P_i&\equiv -\partial_i\varphi(\dot{\bar\phi}+\dot\varphi) \, .
\eea
Now we can solve the constraints \refeq{consmpt} in a similar fashion as when we do perturbation theory in the field fluctuation. The solution is 
\bea
\psi^{(1)}&=&a^2\partial^{-2}\left(3\partial^{-2}\partial_iP_i- L \right) \, \nonumber,\\
\tilde N_i^{(1)}&=&2a^2\partial^{-2}\left(P_i-\partial^{-2}\partial_i \partial_jP_j\right) \, ,\\
N^{(1)}&=&-\frac{1}{2H}\partial^{-2}\partial_iP_i \, .\nonumber
\eea
These solutions to the constraint equations are correct to all orders in the field fluctuations $\varphi$, and first order in $\Mpl^{-2}$.


\subsection{Comoving Gauge}

The solutions of the constraint equations in comoving gauge $ \varphi=0 $, $\zeta \ne 0$ are known to second order in fluctuations $\zeta$ \cite{Arroja:2008ga}. To say that they are cumbersome is an understatement. In the decoupling limit, the full metric in flat gauge is  
\beq
 \dd s^2 = -\dd t^2 + a(t)^2 \dd x^2 + O(\Mpl^{-2}) ,
\eeq
where we discarded the graviton. In this subsection, we show that most of the terms in the lapse and shift in $\zeta \ne 0 $ gauge are captured by the coordinate transformation that takes one from flat gauge to comoving gauge. At each order in perturbation theory, working out the coordinate transformation between $\varphi$ and $\zeta$ is much simpler than solving the constraint equations explicitly. Incorporating the $\varepsilon$ suppressed corrections might bring back all the original difficulty, but the explicit checks at first and second order should convince the reader that there is some advantage in having this different perspective on the solution to the lapse and shift in $\zeta$ gauge. 

\subsubsection{First Order}

The flat gauge is characterized by
\beq\label{fg}
\phi(\tilde t, \tilde x)=\bar\phi(\tilde t)+\varphi (\tilde t, \tilde x);~~ h_{ij}(\tilde t, \tilde x)=a(\tilde t)^2\delta_{ij},~ N=1, ~N_i=0, 
\eeq
while the comoving gauge is
\beq\label{cg}
\phi(t,x)=\bar\phi(t);~~~~~~ h_{ij}(t,x)=e^{2\zeta(x,t)}a(t)^2\delta_{ij} ~.
\eeq
To first order in fluctuations, all we need is a time reparametrization to go from \refeq{fg} to \refeq{cg}. The explicit coordinate transformation is
\beq\label{con}
\tilde t = t + T(t,x) , ~\tilde x^i = x^i .
\eeq
Taylor expanding the field $\phi(\tilde t, \tilde x)$ to first order in fluctuations, we obtain
\beq
\bar\phi(t+T)+\varphi(t+T, x) = \phi (t) \Rightarrow T_{(1)} = - \frac{\varphi(t,x)}{\dot{\bar\phi}(t)} .
\eeq
From this expression it is quite easy to obtain the formula that relates $\zeta$ to $\varphi$. It follows from expanding the scale factor using the new time coordinate,
\beq
a(\tilde t)^2 \approx a(t)^2 (1+2 H T_{(1)}+\cdots) \Rightarrow \zeta= H T_{(1)}+{\cal O}(\varphi^2)
\eeq
With these formulas in hand, we can perform the change of coordinates \refeq{con} to obtain expressions for the new lapse and shift functions, in comoving gauge. By applying the standard diffeomorphism rule for the metric,
\beq
g_{\mu\nu}(t,x)=\frac{\partial \tilde x^\alpha}{\partial x^\mu}\frac{\partial \tilde x^\beta}{\partial x^\nu}\, \widetilde{g}_{\alpha\beta}(\tilde t,\tilde x) \, ,
\eeq
and comparing to the ADM form of the metric \refeq{admmet}, we obtain
\bea
&g_{00}=-(\partial_0 T+1)^2 = -1-2\partial_0 T_{(1)} + {\cal O}(\varphi^2) \Rightarrow N=1+\displaystyle\frac{\dot\zeta}{H}+{\cal O}(\zeta^2)\, ,\\
&g_{0i}=-(\partial_0T+1)\partial_iT=-\partial_i T_{(1)}+{\cal O}(\varphi^2) \Rightarrow N_i = -\displaystyle\frac{\partial_i\zeta}{H}+{\cal O}(\zeta^2) \, .
\eea
Notice that we discarded terms with $\dot H$, as this is the approximation we use in the decoupling limit. We obtain the correct result for the lapse function and for the shift vector, up to a term of order $\varepsilon$ in the shift \cite{Maldacena:2002vr}. It is useful to introduce some notation from Maldacena's paper,
\beq
\frac{\dot\zeta}{H} \equiv \alpha,~~-\frac{\zeta}{H} \equiv \psi \, .
\eeq
This notation will be useful when we discuss the solutions of the constraint equations to second order in fluctuations. Notice that, although $\alpha$ receives no slow-roll corrections, $\psi$ receives a slow roll correction not captured by the decoupling limit. Nonetheless, it is quite remarkable that the solutions to the constraint equations are somewhat trivial in flat gauge, and become fairly complicated in comoving gauge. To obtain the full solution to the linear constraint equations, one needs to incorporate the $\Mpl$ suppressed terms for the metric in flat gauge, which will give the simple substitution
\beq\label{psialpha}
\psi \to \psi + a^2\, \varepsilon\, \partial^{-2} \dot \zeta .
\eeq

The $\varepsilon$ term in the shift is nonlocal in position space, and one can wonder if, in the decoupling limit, one is only able to capture the local pieces in the lapse and shift. That is not the case, once one goes to quadratic order in $\zeta$. In fact, most of the terms in the exact solution \cite{Arroja:2008ga} can be obtained by the change of coordinates from the flat gauge to comoving gauge. This simplicity suggests an alternative method of finding the constraint solutions in comoving gauge. Namely, one considers the solutions to the constraint equations in flat gauge, and determines the coordinate transformation that takes one from flat to comoving gauge. Perhaps one indication that this route might be more economical is that the quadratic change of coordinates was calculated by Maldacena in 2002, and the solutions to the constraint equations to quadratic order were worked out a few years later by Arroja and Koyama, in 2008. 


\subsubsection{Second Order} \label{a:secord}

We proceed in similar fashion as for the first order calculation. The first step, that of finding the change of coordinates, was worked out in \cite{Maldacena:2002vr}. We can simplify a little bit the equations there, as we are in the decoupling limit, and thus have no tensor fluctuations. As a first step, we find the change in the time coordinate that makes the scalar field profile constant at fixed time; namely, we solve the equation
\beq\label{Tsec}
\bar\phi(t+T)+\varphi(t+T)=\phi(t) \Rightarrow T= - \displaystyle\frac{\varphi}{\dot{\bar\phi}}-\frac{\ddot{\bar\phi}}{2\dot{\bar\phi}^3}\varphi^2+\frac{\varphi\dot\varphi}{\dot{\bar\phi}^2} \, .
\eeq
If we proceed to change coordinates, we find that the metric is not in diagonal form in the new coordinate system,
\beq
g_{ij}(t,\tilde x)= - \partial_i T \partial_j T + a(\tilde t)^2 \delta_{ij} \, .
\eeq 
To bring the spatial metric to diagonal form, we must change the spatial coordinates, 
\beq
\tilde t = t+T, \, \tilde x^i = x^i+\xi^i \, ,
\eeq
where $\xi^i$ is, to leading order, quadratic in fluctuations. The calculation is very similar to the one in Appendix A of \cite{Maldacena:2002vr}; the only simplification is in the fact that we neglect the tensor fluctuations. Imposing that the new metric is of diagonal form ($3 g_{ij}- \delta_{ij} g_{aa} =0$), and using the Helmholtz decomposition $\xi^i=\partial_i {\cal A} + \tilde \xi_i$, we obtain
\bea
&{\cal A}&=\frac{3}{4}a(t)^{-2} \partial^{-4}\partial_a \partial_b \left((\partial_a T \partial_b T)-\frac{1}{3} \partial_a^2(\partial_b T)^2 \right)  \, ,\\
& \tilde \xi_i&=a(t)^{-2} \partial^{-2} \left(\partial_a(\partial_a T \partial_i T)-\partial_i \partial^{-2}\partial_a\partial_b\left(\partial_a T\partial_b T\right) \right) \, .
\eea
After finding this change of coordinates, we bring the metric to diagonal form,
\beq
g_{ij} = \left(-\frac{(\partial T)^2}{3}+a(t+T)^2+\frac{2}{3} a(t)^2 \partial^2 {\cal A}\right) \delta_{ij} \, ,
\eeq
and identify $\zeta$, to second order in perturbations, as

\beq
\zeta= H T +\frac{1}{4 a(t)^2}\left(\partial_a \partial_b \left(\partial_a T \partial_b T)-\partial_a^2(\partial_b T)^2 \right)  \right)+\cdots
\eeq

Notice that the first term proportional to $T$ has pieces linear and quadratic in fluctuations (see \refeq{Tsec}). It is useful to solve for $T$ in terms of $\zeta$,
\beq
T=-\psi-\frac{1}{4 a^2 H^2}\left(\partial^{-2}\partial_a\partial_b\left(\partial_a \psi \partial_b\psi\right)-\left(\partial_a\psi\right)^2\right) \,  ,
\eeq
with $\psi$ given by \refeq{psialpha}.

The final step is to analyze the transformed expressions for $g_{00}$ and $g_{0i}$ and read off the quadratic solutions for the lapse and the shift. The analysis is straightforward but tedious, so we state the main result without going through the details. For the metric components $g_{00}$ and $g_{0i}$, we find
\bea
&g_{00}&=-(1+\partial_0T)^2\, , \\
&g_{0i}&=-\partial_iT(1+\partial_0T)+a(t)^2\partial_0\xi_i\, .
\eea
From these expressions we can read off $N_i$ and $N$ to quadratic order in $\zeta$. Arroja and Koyama \cite{Arroja:2008ga} use the standard notation for the quadratic corrections to the lapse and shift,
\bea
&N&=1+\alpha+\alpha^{(2)}\, ,\\
&N_i&=\partial_i\psi+\widetilde N_i^{(2)}+\partial_i\psi^{(2)},~ \partial_i \widetilde N_i^{(2)} = 0\, .
\eea
 The result exact in slow-roll is presented in equations (29-32) in \cite{Arroja:2008ga}; we present the slightly simplified formulas that arise in the decoupling limit,
\bea
&\alpha^{(2)}&=\frac{\partial^{-2}\partial_iF_i}{2H}\, ,\nonumber\\
&\widetilde N_i^{(2)}&=2 a(t)^2\left[\partial^{-4}\partial_i\partial_j F_j-\partial^{-2}F_i \right]\, ,\\
&\partial^{2}\psi^{(2)}&=\frac{H}{2}\left(\partial_i \psi \right)^2+\alpha \partial^2\psi-\frac{1}{4 a^2 H}\left[\left(\partial_i\partial_j\psi\right)^2-\left(\partial^2\psi\right)^2\right] -\frac{3 }{2a^2}\partial^{-2}\partial_iF_i\, , \nonumber\\
&F_i&\equiv-a(t)^{-2}\left(\partial_j\alpha\partial_j\partial_i\psi-\partial_i\alpha\partial^2\psi-H \partial_j\left(\partial_i\psi \partial_j \psi\right)\right)\, ,\nonumber
\eea
with $\alpha$ and $\psi$ given by \refeq{psialpha}. It is rather remarkable that most of the structure of the constraint equation solutions is captured by a change of coordinates from the flat metric. Once again, this change of coordinates might be a more practical way to obtain solutions of the constraint equations in comoving gauge.

\subsection{To What Order?}
\label{constrtoorder}

In order to find the cubic action for the gravitational fluctuations, Maldacena noticed that one needs to solve the constraint equations only to first order. In this section, we show that, with the $n^{\rm th}$ order solution to the constraint equation, one can obtain the $(2n+1)^{\rm th}$ order action.

We need to find the Lagrangian ${\cal L}(h,\bar N,\bar N_i)$ with the constraints $\bar N(h)$ and $\bar N_i(h)$ evaluated at their respective saddle points (we suppress the spatial metric indices). In other words, we find $\bar N$ and $\bar N_i$ by solving the equations

\bea\label{ce1}
&\displaystyle\frac{\delta {\cal L}}{\delta N}(h,\bar N,\bar N_i)&\equiv \frac{\partial {\cal L}}{\partial N}-\partial_i\frac{\partial {\cal L}}{\partial (\partial_i N)}=0, \\ \,&\displaystyle \frac{\delta {\cal L}}{\delta N_i}(h,\bar N,\bar N_i)&\equiv\frac{\partial {\cal L}}{\partial N_i}-\partial_j\frac{\partial {\cal L}}{\partial (\partial_j N_i)}=0.
\eea

Solving the constraint equations to $n^{\rm th}$ order in perturbation theory means that 

\beq
\bar N = N^{(n)}+ O(\zeta^{n+1}), \, \bar N_i = N^{(n)}_i+O(\zeta^{n+1})\, .
\eeq
Now, if we write the action in terms of $N^{(n)}$ and $N_i^{(n)}$, we obtain 
\bea \label{ce2}
& \displaystyle\int {\cal L} (h, \bar N, \bar N_i)=& \int {\cal L} (h, N^{(n)}, \bar N_i^{(n)})+\\&&+\int \left(\bar N - N^{(n)}\right)\frac{\delta {\cal L}}{\delta N} (h, N^{(n)},N_i^{(n)})+\int \left(\bar N_i - N_i^{(n)}\right)\frac{\delta {\cal L}}{\delta N_i} (h, N^{(n)},N_i^{(n)}) + \nonumber \\ &&+\int \partial_j\left(\left(\bar N - N^{(n)}\right)\frac{\partial {\cal L}}{\partial (\partial_j N)} (h, N^{(n)},N_i^{(n)}) \right)+\, \nonumber \\&&+\int \partial_j\left(\left(\bar N_i - N_i^{(n)}\right)\frac{\partial {\cal L}}{\partial (\partial_j N_i)} (h, N^{(n)},N_i^{(n)}) \right)\,  \nonumber\\ &&+ O\left(\left(\bar N - N^{(n)}\right)^2,\left(\bar N_i - N_i^{(n)}\right)^2\right)\, , \nonumber
\eea
where $(\delta/\delta N, \delta/\delta N_i)$ are variational derivatives. This is a Taylor expansion of the full action using the approximate solution to the constraint equation. Let us unpack each line of the right hand side of \refeq{ce2}. In the first line we are evaluating the Lagrangian on the solutions $(N^{(n)},N_i^{(n)})$. The second line plus the third and fourth lines give the linear Taylor expansion of the action. We reorganize the terms in total derivatives, which are in the third and fourth lines, plus a term proportional to the constraint equations themselves, in the second line. Notice that this manipulation is crucial, as there are terms in the Lagrangian that contain the extrinsic curvature $K_{ij} \sim \nabla_{(i} N_{j)}$. 

The terms in the second line are precisely the constraint equations \refeq{ce1}, which are solved by $(N^{(n)}, N^{(n)}_i)$ to the $n^{\rm th}$ order. This means that the error made in the Taylor expansion \refeq{ce2} is of order $2(n+1)$. One $n+1$ comes from the error in $\bar N- N^{(n)}$, and the other $n+1$ from the error in $\delta {\cal L}/\delta N$. Notice that the error is of the same order as the term in the fifth line of \refeq{ce2}. Finally, the terms in the third and fourth lines of \refeq{ce2} are total derivatives, and can thus be safely discarded.

In summary, by using the $n^{\rm th}$ order solution to the constraint equation, we make an error of order $2(n+1)$ in fluctuations. This implies that we can trust our perturbative expansion of the action to order $2n+1$.


\section{From potential to Hubble parameters}
\label{PotinSRpar}

By taking successive time derivatives of
\be
V(\bar{\phi}) = \Mpl^2H^2(3-\varepsilon)\,,
\label{V}
\ee
and using the chain rule $ \partial_{t}=\dot \phi\partial_{\phi} $, we find
\be\label{Vp}
V'(\bar{\phi})&=&\Mpl H^2\left[- \sqrt{\frac{\varepsilon}{2}}\eta-3 \sqrt{2\varepsilon}+\sqrt{2 \varepsilon}\varepsilon\right]\,,\\
V''(\bar{\phi})&=& H^2 \left[-\frac{3}{2}\eta+\frac{5}{2}\varepsilon\eta - \frac{1}{4}\eta^2-\frac{1}{2}\frac{\dot{\eta}}{H} - 2\varepsilon^2+6\varepsilon\right]\,,
\label{Vpp}\\
V'''(\bar{\phi})&=& \frac{H^2}{\sqrt{2\varepsilon}\Mpl}\left[ - \frac{3}{2}\frac{\dot{\eta}}{H}-\frac{\ddot{\eta}}{2H^2}-\frac{\eta\dot{\eta}}{2H}+9 \varepsilon\eta+3\frac{\varepsilon\dot{\eta}}{H}+3\varepsilon\eta^2-9\varepsilon^2\eta + 4\varepsilon^3-12\varepsilon^2\right]\,.\label{Vppp}
\ee

 
\section{The Power Spectrum with a Time Dependent Mass }\label{a:gfct}

In this appendix, after reviewing some standard facts about canonical quantization of a free scalar field in de Sitter space, we compute the two point function of inflaton fluctuations $ \varphi_{s} $ in the presence of a time dependent mass, \refeq{tdm}, induced by a long inflaton mode $ \varphi_{l} $. This is then used in \refss{ssec:soft} to calculate the squeezed limit of the bispectrum. We are able to carry on the calculation in three cases. In \refss{ssec:intdim} we discuss the cases in which $ V''/H^{2}=0 $ and $ V''/H^{2}=2 $. For these cases, $ \nu $ in \refeq{modefuncsol} is half integer, $ \nu=0 $ and $ \nu=1/2 $, and therefore the Hankel function reduces to simple functions, which we can integrate analytically. These correspond to dimensions $ \Delta_{-}=0 $ and $ \Delta_{-}=1 $ respectively, so we refer to them as integer dimensions. In \refss{ssec:intdim2} we discuss the case of small mass, $ V''\ll H^{2} $, where we work to leading order in $ V'' $.


\subsection{Quadratic Action}\label{a:ff}
In this section, we review some standard facts about canonical quantization of fluctuations in an inflating spacetime. As we saw in the \refapp{a:A}, to all orders in fluctuations $\varphi$, the constraint equations are trivial to leading order in $\Mpl$. Then, the quadratic scalar action is simply that of a scalar field in de Sitter spacetime

\beq
S_{2}=\frac{1}{2}\int \dd^{4}x \, a^{3} \,\left[  \dot \varphi^{2}-\frac{(\partial_i \varphi)^{2}}{a^{2}}-V''(\bar\phi)\varphi^{2}\right]\, .\\
\label{S2}
\eeq
To calculate the $\varphi$ power spectrum, we quantize the scalar perturbation as
\beq
\varphi(\tau,\textbf{k}) = u_k (\tau) \hat{a}_{\textbf{k}}+u^{\ast}_k(\tau)\hat{a}^{\dagger}_{\textbf{k}}\,,
\label{Fdecomp}
\eeq
where $\hat{a}_{\textbf{k}}$ and $\hat{a}^{\dagger}_{\textbf{k}}$ satisfy the usual commutation relation, 
\beq
\left[\hat{a}(\textbf{k}),\hat{a}^{\dagger}(\textbf{k}') \right]= (2\pi)^3 \delta^{(3)}(\textbf{k}+\textbf{k}').
\eeq
The equations of motion for the mode functions are
\beq
u_k'' - \frac{2}{\tau}u_k'+\left(k^2  + \frac{V''}{H^2 \tau^2}\right)u_k = 0\,,
\label{eomphi}
\eeq
where we suppressed the $\bar \phi $ dependence of the \textit{effective mass of perturbations} $ V''(\bar\phi) $. In terms of slow-roll parameters,
\beq
V''(\bar{\phi})= H^2 \left[-\frac{3}{2}\eta+\frac{5}{2}\varepsilon\eta - \frac{1}{4}\eta^2-\frac{1}{2}\frac{\dot{\eta}}{H} - 2\varepsilon^2+6\varepsilon\right]\,.
\eeq
Therefore, the slow-roll conditions imply 
\be\label{etaVeta}
\eta_{V}\equiv \Mpl^{2}\frac{V''}{V}\simeq -\frac{\eta}{2}\ll1\,,
\ee
and so $V''\ll H^2$. The mode function with Bunch-Davies initial conditions is
\beq
u_k(\tau) = - e^{-i\frac{\pi}{2}(\nu+ \frac{1}{2})}\frac{\sqrt{\pi}}{2}H (-\tau)^{\frac{3}{2}}\text{H}_{\nu}^{(1)}(-k \tau),
\label{modefuncsolapp}
\eeq
where 
\be\label{nu}
\nu \equiv \sqrt{9/4 - V''/H^2}\,,
\ee
and $\text{H}_{\nu}^{(1)}$ is the Hankel function of the first kind. For $V''/H^2 > 9/4$, $ \nu $ is pure imaginary $\nu=i|\nu|$. Also, note that the Hankel functions can be written in terms of simple functions only when $\nu$ is half-integer, for example
\be\label{dSmodefct}
u_k(\tau) &=& \frac{H}{\sqrt{2k^{3}}} \left(  1+ik\tau\right)e^{-ik\tau}\,, \quad (\nu=3/2)\,,\\
u_k(\tau) &=& -\frac{H\tau}{\sqrt{2k}} e^{-ik\tau}\,,\quad\quad\quad\quad \quad (\nu=1/2)\,.
\ee
At late times $-k\tau \ll 1$, the asymptotic behavior of \refeq{modefuncsolapp} is mass dependent, 
\beq
u_k(\tau) \approx i \frac{2^{\nu-1} \Gamma(\nu) }{\sqrt{\pi}}H \frac{ (-\tau)^{3/2}}{(-k\tau)^{\nu}}\,, \label{modefuncnu} 
\eeq
As mentioned before, the relevant solution for inflation is $|V''|/H^2\ll1$. However, as we will see later on, the more general but physically less relevant solutions provide a non-trivial check of the correctness of our approach in \refsec{s:softth}.

 
\subsection{Integer Dimensions}\label{ssec:intdim} 

The equation of motion for $\varphi_s$ in the presence of long and therefore approximately spatially constant $ \varphi_{l} $ is
\beq
\partial_{\tau}^2\varphi(\tau)-\frac{2}{\tau}\partial_{\tau}\varphi(\tau)+\left(k^2 + \frac{V''}{\tau^2 H^2} \right)\varphi(\tau)=- \frac{V'''(\bar{\phi})}{H^2 \tau^2}\hat{\varphi}_{l}(k_l)\tau^{3/2-\nu} \varphi(\tau)\,,
\label{EMS}
\eeq
where $\hat{\varphi}_l(k_l)$ is the time independent part of \refeq{modefuncnu} and $k_l$ is the momentum of the long mode and for notational convenience, we dropped the subscript $s$ of the short modes. Notice that the effective mass of $ \varphi_{s} $ is time dependent. Also, note that we used the late time behavior of the mode functions, i.e. we used \refeq{modefuncnu}. \refeq{EMS} resembles Bessel's equation, which, in the presence of a generic time dependent term, cannot be solved exactly. On the other hand, we are only interested in the linear correction in $ \varphi_{l} $, so we can just solve it perturbatively using the Green's function.

Let us first start with a pedagogical review of Green's functional methods. A general inhomogeneous differential equation has the form
\beq
\deq(\tau) \varphi(\tau) = f(\tau),
\label{deq1}
\eeq
here $\deq$ is a differential equation in terms of $\tau$ and $f(\tau)$ is some function. To find a solution for $\varphi(\tau)$, we split up the solution for $\varphi$ into a homogeneous and a particular solution
\beq
\varphi(\tau)=\varphi_H(\tau)+\varphi_p(\tau)\,,
\eeq
where $\varphi_H$ solves \refeq{deq1} for $f(\tau)= 0$. We consider a particular solution of the form
\beq
\varphi_p(\tau) = \int_a^b d\tau'~ G(\tau,\tau')f(\tau')\,,
\label{parsol}
\eeq
where $G(\tau,\tau')$ is the Green's function and $a$ and $b$ define the domain for which the particular solution of $\varphi$ should be valid. Of course, when we act on this particular solution with the differential operator $\deq$, we should obtain $f(\tau)$ again,
\beq
 \int_a^b d\tau'~ \deq(\tau) G(\tau,\tau')f(\tau') = f(\tau)\,.
\label{parsol}
\eeq
Therefore, the Green's function should satisfy the following relation 
\beq
\deq(\tau) G(\tau,\tau') = \delta(\tau-\tau')\,,
\eeq
here $\delta(\tau - \tau')$ is the Dirac delta function. We now solve for $\varphi_{p}$ iteratively. To do this, we split $\varphi_{p}$ as
\beq
\varphi_{p}(\tau,k) = \sum_{n=1}^{\infty} \varphi^{(n)}\,,
\label{pephi}
\eeq
where the superscript $(n)$ indicates a correction of the order $\left(V''/H^{2}\right)^n$. Substituting \refeq{pephi} into \refeq{EMS}, we obtain  the following differential equation for the $(n+1)$-th order corrections to the mode functions
\beq
\partial_{\tau}^2\varphi^{(n+1)}(\tau)-\frac{2}{\tau}\partial_{\tau}\varphi^{(n+1)}+\left(k^2 + \frac{V''}{\tau^2 H^2} \right)\varphi^{(n+1)}= -\frac{V'''(\bar{\phi})}{H^2 \tau^2}\hat{\varphi}_{l}\tau^{3/2-\nu} \varphi^{(n)}\,.
\label{deq3}
\eeq 
The Green's function of \refeq{EMS} is
\beq
\left( \Box - V'' \right)G(\tau,\tau') = \frac{1}{\sqrt{-g}}\delta(\tau - \tau')\,,
\eeq
Working out the D'Alembertian explicitly, we obtain
\beq
\partial_{\tau}^2G(\tau,\tau')-\frac{2}{\tau}\partial_{\tau}G(\tau,\tau')+\left(k^2 + \frac{V''}{\tau^2 H^2} \right)G(\tau,\tau')=\tau^2 H^2 \delta(\tau - \tau')\,.
\label{deq4}
\eeq
In the language of \refeq{deq1}, the function $f(\tau)$ for the linear order correction to the mode functions is
\beq
f(\tau) \equiv - \frac{V'''(\phi_0)}{H^2 \tau^2}\hat{\varphi}_{l}\tau^{3/2-\nu} \varphi^{(0)}(\tau)\,.
\eeq 
As mentioned before, there are two cases for $\nu$ for which we can do exact calculations, these corresponded  to the massless case $V''/H^2=0$ and to the massive case $V''/H^2 = 2$. The solutions of the Green's function for both these cases are respectively
\bea
&&G_{3/2}(\tau,\tau')=\frac{H^2}{k^3}\left[(1+k^2 \tau \tau')\sin\left(k(\tau-\tau')\right)-k(\tau-\tau')\cos\left(k(\tau-\tau')\right) \right] \Theta(\tau - \tau')\,,\label{G32}\\
&&G_{1/2}(\tau,\tau')=\frac{H^2}{k}\tau \tau'\sin\left(k(\tau-\tau')\right)\Theta(\tau - \tau')\,.\label{G12}
\eea
The linear corrections to the particular solutions that can be obtained from \refeq{G32} and \refeq{G12} are then 
\bea
&&\varphi_{p,3/2}^{(1)}(\tau,k_s)=\frac{V'''(\bar{\phi}) \hat{\varphi}_l(k_l)e^{-i k_s \tau}}{3\sqrt{2 k_s^3}H}\left[-2+i e^{2 i k_s \tau}(i+ k_s \tau)\Gamma(0,2 i k_s \tau) \right]\,,\label{parsol32}\\
&&\varphi_{p,1/2}^{(1)}(\tau,k_s)=\frac{i V'''(\bar{\phi})\hat{\varphi}_{l}(k_l) \tau e^{-i k_s \tau}}{2 \sqrt{2 k_s^3}}\left[e^{2i k_s \tau}\text{ExpEi}(-2 i k_s \tau)-2 \log({k_l\tau)} \right]\,.
\label{parsol12}
\eea
where $ \Gamma(z,a) $ is the incomplete gamma function

 
\subsection{Small Masses}\label{ssec:intdim2}

We have been able to compute the squeezed limit bispectrum also at leading non-vanishing order in $ V''\propto \eta $. The final expression is complicated and we have not been able to use conformal symmetry to extend it to the full shape, away from the squeezed limit. We briefly report this result below.

Since we are interested in scalar fields for which the mass is very small, $V''/H^2\ll1 $, we can perturbatively solve the linearized equation of motion. We consider
\beq
\partial_{\tau}^2\varphi(\tau)-\frac{2}{\tau}\partial_{\tau}\varphi(\tau)+k^2 \varphi(\tau)= \frac{1}{H^2 \tau^2}\left(- V'' - V'''(\bar{\phi})\hat{\varphi}_{l}\tau^{\frac{V''}{3H^2}}\right) \varphi(\tau)\,.
\label{EMS2}
\eeq
The particular solution is long an uninspiring. We used it in the consistency relation \refeq{squeezed5} to derive the following squeezed limit of the bispectrum
\beq
\begin{split}
&\langle \varphi(\textbf{k}_s,\tau_*)\varphi(\textbf{k}_s,\tau_*)\varphi(\textbf{k}_l,\tau_*)  \rangle'=	\frac{H^2V'''}{6}\frac{(-2 + \gamma_E + \log{(-2 k_s \tau_*)})}{k_s^3 k_l^3}+\\
&+\frac{V'' V'''}{k_l^3 k_s^3}\left[\frac{1}{9}[-2 + \gamma_E +\log\left(-2 k_s \tau_* \right)]\log\left(\frac{k_l}{k_s} \right) -\frac{7}{108}[-2 + \gamma_E +\log\left(-2 k_s \tau_* \right)] +\right.\\
&\left.+\frac{1}{6}[-2 + \gamma_E +\log\left(-2 k_s \tau_* \right)][-2 + \gamma_E +\log\left(-2 k_s \tau_* \right)]-\frac{25}{432}\pi^2 \right]+\O\left(m_{0}^{4},\left( V''' \right)^{2} \right)\,.
\label{lightmassB}
\end{split}
\eeq
Part of these logs can be re-summed into exponentials of $ k $ using dilation invariance to guess the correct form. The remaining logs represent the dilation anomaly discussion below \refeq{D3pt}.


\section{Conversion between $ \zeta $ and $ \varphi $} \label{a:last}

In this appendix we perform the gauge transformation from flat to comoving gauge, using the $\delta N$-formalism \cite{Sasaki:1995aw}. In the $\delta N$ formalism, we discard terms that vanish outside of the Hubble radius. Our derivation is rather standard, but in \refsec{sec:MMM} we provide a derivation of the $\delta N$ formalism from the wavefunctional of the universe point of view. We comment on the more general method of performing the full gauge transformation from $\varphi$ to $\zeta$ in \refapp{a:A}. 

The key observation in the $\delta N$-formalism is to relate a long $\zeta$ fluctuation to a local perturbation to the scale factor of the universe. This, in turn, is equal to a perturbation of the number of e-folds, $N$, which arises from perturbing the initial scalar field $\phi$ in the spatially flat gauge. This observation suggests us to consider a local scale factor
\beq
a(\textbf{x},t) \equiv a(t)e^{\zeta(\textbf{x},t)}\,.
\eeq
The number of e-folds from an arbitrary initial time $t_0$ to the time of Hubble exit $t_*$ is
\beq
N = \int_{t_0}^{t_*} dt'~H(t') = \ln{\left( \frac{a(t_*)}{a(t_0)}\right)} =H( t_*-t_0)\,.
\label{Ndef}
\eeq
If we define $N(\textbf{x},t)$ as the number of e-folds from a fixed flat slice to a comoving curvature slice at time $t$, then 
\beq
\zeta(\textbf{x},t) = \delta N(\textbf{x},t)\,.
\eeq
To relate $\zeta(\textbf{x},t)$ to the superHubble inflaton perturbations $\varphi(\textbf{x},t)$, we proceed as follows. We choose a spatially flat initial time-slice on which there are no scalar fluctuations in the metric, but only fluctuations in the matter fields $\phi(\textbf{x},t)=\bar{\phi}(t)+\varphi(\textbf{x},t)$. We then choose a final time slice to be a comoving curvature slice, where the scalar fluctuations are eaten by the metric. To go from one slice to the other, we separately evolve the unperturbed and the perturbed fields classically to the final slice. The difference between the two results is then equal to the difference in the number of e-folds,
\beq
\zeta = \delta N = N(\bar{\phi}) - N(\bar{\phi} + \varphi) \,.
\label{zetadeltaN}
\eeq
Expanding \refeq{zetadeltaN} around $\bar{\phi}$ we  obtain an expression for $\zeta(\textbf{x},t)$ in terms of the scalar fluctuations $\varphi$ and derivatives of $N$ defined on the initial slice,
\beq
\zeta \approx -	N'(\bar{\phi})\varphi - \frac{1}{2}N''(\bar{\phi})\varphi^2+...\,,
\eeq
here we assumed that $\varphi \ll \bar{\phi}$ and the primes denote derivatives with respect to $\phi$. To convert the $\varphi$ two point function to the $\zeta$ power spectrum, a first order relation in $\varphi$ is sufficient
\be
\left.\frac{\partial N}{\partial \phi}\right|_{\bar\phi} = \frac{H}{\dot{\bar\phi}}= \frac{1}{\Mpl \sqrt{2\e}},
\ee
where we used $\dot {\bar\phi}= \sqrt{2\e} \Mpl H$. To determine the $\zeta$ bispectrum, we also need $\zeta$ to $O(\varphi^2)$
\be
\Mpl^2\left.\frac{\partial^2 N}{\partial \phi^2}\right|_{\bar\phi} = \frac{1}{\sqrt{2\e}H}\frac{d}{dt} \frac{1}{\sqrt{2\e}}
=- \frac{\eta}{4\e}.
\ee
So to second order we have
\beq
\zeta = -\frac{1}{\sqrt{2\e}\Mpl}\varphi + \frac{\eta}{8\e\Mpl^2}\varphi^2.
\label{zetadeltaN2}
\eeq
This is equivalent to the second order transformation found in \cite{Maldacena:2002vr}:
\be
\zeta = H T +\frac{1}{2} \dot H T^2
\ee
with $T$ given by 
\be
T = -\frac{1}{\sqrt{2\e}\Mpl}\varphi - \left(\frac{\eta}{8\e\Mpl^2}-\frac{1}{4}\right)\varphi^2 
+\frac{1}{2\e\Mpl^2 H}\varphi \dot{\varphi},
\ee
where at linear order $\dot\varphi \approx \frac{1}{2}\eta H\varphi$ up to an exponentially decaying term.

 
\subsection{The Bispectrum the Third Way}\label{ssec:btw}

Here, we discuss the third natural choice for the conversion of the $ \varphi $ bispectrum into the bispectrum of $ \zeta $. In analogy to what we did for the power spectrum in \refss{ss:ps}, we can choose to convert to $ \zeta $ at Hubble crossing
\beq
-k \tau_{\text{H.c.}}=1\,.
\eeq
However, since the wavenumebers in the bispectrum are generally different, they exit the Hubble radius at different times. To ensure that all modes are frozen by the time we change from $\varphi$ to $\zeta$, we choose $\tau_{\text{H.c.}}$ to be the time when the shortest wavelength mode has become Hubble-sized. Without loss of generality (due to Bose symmetry) we choose $ k_{1}\leq k_{2}\leq k_{3} $ and so $\tau_{\text{H.c.}}= -1/k_3$. 

Converting the massless $\varphi$ bispectrum, \refeq{B32}, to the $ \zeta $ bispectrum, one obtains
\be\label{varphiresult3}
\langle \zeta_{\k_{1}} \zeta_{\k_{2}} \zeta_{\k_{3}}\rangle &=& \frac{H^4}{16 \e^2\Mpl^4}\frac{\eta}{2} \Bigg|_{\tau_{\text{H.c.}}}\frac{k_1^3 + k_2^3+k_3^3}{k_1^3 k_2^3k_3^3} +  \frac{H^4}{16 \e^2\Mpl^4}\frac{\dot{\eta}}{2H}\frac{1}{k_1^3k_2^3k_3^3}\Bigg|_{\tau_{\text{H.c.}}}\\
&&\times \left[(-1+\gamma_E+ \log{(-K \tau_{\text{H.c.}})})\sum_{i=1}^3 k_i^3-\sum_{i\neq j}k_i^2 k_j +  k_1 k_2 k_3\right]\,.\nonumber
\ee


\newpage
\bibliographystyle{utphys}
\bibliography{ConfLimit-Refs}

\end{document}